\documentclass[preprint, 3p]{elsarticle}
\usepackage{float}
\usepackage{xcolor}
\usepackage{graphicx}
\usepackage{subfigure}
\usepackage{nomencl}
\usepackage{enumerate}
\usepackage{verbatim}
\usepackage{amssymb}
\usepackage{amsmath}
\usepackage{natbib}
\usepackage{placeins}
\usepackage{epstopdf}
\usepackage{rotating} %{realboxes}
\graphicspath{{Figures_eps/}}

 \makeatletter

\newcommand{\Rmnum}[1]{\expandafter\@slowromancap\romannumeral #1@}
\makeatother

\title{The discrete Green's function paradigm for two-way coupled Euler-Lagrange simulation}

\author[]{J. A. K. Horwitz$^{1,2}$\corref{cor1}}
\author[]{G. Iaccarino$^{2}$}
\author[]{J. K. Eaton$^{2}$}
\author[]{A. Mani$^{2}$}

\cortext[cor1]{Corresponding Author, email: horwitz1@stanford.edu}

\address{$^1$Lawrence Livermore National Laboratory, P.O. Box 808, Livermore, CA 94550-0808\\
$^2$Stanford University, Department of Mechanical Engineering} %,\\

\journal{Arxiv}

\begin{document}
%\maketitle
\begin{abstract}

We outline a methodology for the simulation of particle-laden flows whereby the %momentum, and/or other quantities of the 
dispersed and fluid phases are two-way coupled. The drag force which couples fluid and particle momentum depends on the undisturbed fluid velocity at the particle location, and this latter quantity requires modelling. %In the context of momentum, the coupling is owing to a force that appears in the respective particle and fluid momentum equations. %In the limit of low particle Reynolds number, the force is well-described by Stokes drag. 
%The drag force, be it Stokesian or otherwise, depends on the undisturbed fluid velocity at the particle location. 
%This quantity is not directly computable real-time in numerical simulations but can be approximated using different strategies. 
%In this work, 
We demonstrate that the undisturbed fluid velocity, in the low particle Reynolds number limit, can be related exactly to the discrete Green's function of the discrete Stokes equations. %owing to a discrete force density. 
%The discrete Green's function encodes information relating to numerical discretization and boundary conditions. 
The method is general in that it can be extended to other partial differential equations which may be associated with particle-laden flows, such as the thermal energy equation or Maxwell's equations. In this work, we demonstrate the method of discrete Green's functions by obtaining these functions for the Navier-Stokes equations at low {\color{black}particle} Reynolds number in a two-plane channel geometry. We perform verification at low and finite particle Reynolds number for the case of a point-particle settling under gravity parallel to a plane wall, for different wall normal separations. In comparing to other point-particle schemes the discrete Green's function approach is the most robust at low particle Reynolds number, accurate at all wall-normal separations and is the most accurate in the near wall region at finite Reynolds number.  
We discuss how the accuracy away from the wall at finite Reynolds number could be improved by appealing to Oseen-like discrete Green's functions. Finally we demonstrate that the discrete Green's function approach can have important implications on statistics of particle-laden turbulent channel flow.
%At finite particle Reynolds number, the discrete Green's function approach, formulated based on the Stokes equations becomes less accurate in the bulk of the channel, but is still the most accurate near the wall.
\end{abstract}

\maketitle

\section{Introduction}
\label{sec:intro}
Particle-laden flows--owing to the abundance of applications to which they are encountered including sand storms \citep{Kok-2012}, soot emissions \citep{Malmborg_2017}, and fluidized beds \citep{Cocco_2014}--have garnered considerable interest in the scientific community. Particle-laden flows are often studied via Euler-Lagrangian (EL) simulation. In EL methods, partial differential equations (such as mass, momentum, temperature, electric-field) for the fluid or carrier phase are discretized in space while ordinary differential equations (such as position, momentum, temperature, and charge) governing the particle or dispersed phase are solved along Lagrangian trajectories for each particle. EL methods are closed by assuming models for how the particles will interact with the fluid and with each other. One-way coupled EL methods describe models whereby particle quantities (position/momentum etc.) are influenced by the surrounding fluid environment, but fluid quantities (velocity, temperature) feel no effect from the particles. In two-way coupled approaches, particles are influenced by the surrounding fluid fields, and the fluid fields themselves are influenced by the particles. This can happen, for example, when a hot particle is placed in a cold surrounding fluid. A two-way coupled EL approach will model how the particle will cool down and how the fluid will heat up \citep{Horwitz_APS_DFD_2017,Balachandar_2019b}. In four-way coupled EL methods, in addition to the mutual interaction of particles and fluid, explicit models are used to account for the mutual influence between particles. Mutual interactions between particles include collisions which causes a change in momentum \citep{cundall_dem} and charge \citep{Yao_Capecelatro_2018}, depending on the application, of the colliding particles. Lubrication and multi-particle hydrodynamic interactions can also be modelled analytically through mobility functions \citep{kim-karrila-2005} or numerically using approaches like the PIEP method \citep{Akiki-2017}. 

In this work, we are concerned principally with issues that arise in two-way coupled EL methods. To elucidate the challenge, let us consider only hydrodynamic interactions between a fluid and a single rigid spherical particle. When the flow is steady and the particle Reynolds number $Re_{p} = |\mathbf{\tilde{u}}-\mathbf{v_p}|d_p/\nu \ll 1$, the interaction between the fluid and particle is well-described by the Stokes drag force $\mathbf{f_d} = 3\pi\mu d_p (\mathbf{\tilde{u}}-\mathbf{v_p})$. Here, $\mathbf{v_p}$ and $d_p$ are respectively the particle velocity and diameter, $\mu$ and $\nu$ are respectively the fluid dynamic and kinematic viscosity respectively, and $\mathbf{\tilde{u}}$ is the undisturbed fluid velocity centered at the particle position \citep{maxey_riley_1983}. In a two-way coupled EL simulation, this Stokes drag force felt by the particle models the rate of change of momentum deficit in the fluid. The momentum deficit in the fluid is characterized by a disturbance flow created by and typically occurring in the near-field of the particle. This disturbance flow contaminates the surrounding fluid so that the undisturbed state prior to the introduction of the particle is no longer available. In other words, it is not  possible to calculate $\mathbf{\tilde{u}}$ which is needed to compute the drag force, changing the motion of the particle, and the fluid velocity, as the simulation is being performed. This observation has been the subject of considerable research in recent years. 

Within the past five years, several methods have been proposed to estimate the undisturbed fluid velocity for momentum two-way coupled EL methods \citep{gualtieri_etal_2015,Horwitz_2016,ireland_desjardins_2017,Horwitz_2018,Fukada_etal_2018,Esmaily_Horwitz_2018,Balachandar_2019a,Poustis_etal_2019} as well as the undisturbed temperature for thermally two-way coupled EL methods \citep{Horwitz_APS_DFD_2017,Balachandar_2019b}. Each of these methods have shown similar accuracy in the context of laminar verification problems while they have different ranges of applicability. Some of these methods are properly suited for low particle Reynolds number \citep{gualtieri_etal_2015,Horwitz_2016,ireland_desjardins_2017} while others have been extended to finite Reynolds number \citep{Horwitz_2018,Fukada_etal_2018,Esmaily_Horwitz_2018,Balachandar_2019a,Poustis_etal_2019}. While some methods are suited to isotropic grids \citep{Horwitz_2016,Horwitz_2018,Horwitz_APS_DFD_2017}, other methods can be employed on more general grid configurations \citep{gualtieri_etal_2015,ireland_desjardins_2017,Balachandar_2019b,Esmaily_Horwitz_2018,Balachandar_2019a,Poustis_etal_2019}. A summary of these methods is presented in Table \ref{tab:err_bar}. Besides which drag correlation and what type of mesh structure these methods are suited to, one area that has not received attention in the literature is the manner in which the domain and discretization of field equations contributes to how the undisturbed fluid velocity should be estimated. These previous methods implicitly assume the fluid domains have homogeneous or periodic boundary conditions. When the boundary conditions become more complicated, for example when one or more no-slip/no-penetration walls is introduced, these methods are no longer strictly valid. An inherent assumption is that the disturbance flow created by a particle can freely decay. In the language we will adopt in the next section, these methods assume the disturbance flow to be given by the discrete Green's function response in an unbounded domain. At large distances, this will go like $\sim 1/r$, or the asympototic form of the \textit{continuous} Stokeslet or Green's function of the Stokes equations. When a particle moves near a wall, especially at close distances, the disturbance flow the particle creates must be damped by the wall to satisfy the no-slip/no-penetration condition \citep{Blake_Chwang_1973,Liron_Mochon_1975}. None of the aforementioned methods explicitly account for the presence of the wall. While several of the previous methods are suited to anisotropic grids which are common in the simulation of wall-bounded flows, we will show in the verification section that capturing the wall-damping effect is essential to accurately computing the two-way coupling force. We will also demonstrate explicitly how discretization of the field equations (via finite differences or finite elements e.g.) is connected to the discrete disturbance flow and ultimately the computation of the undisturbed fluid velocity. {\color{black} It is worth mentioning that the exact regularized point-particle method (ERPP) \citep{gualtieri_etal_2015} was recently extended to the wall-bounded flow regime \citep{Battista_2019}. This approach is particularly attractive in its incorporation of unsteady effects, however, ERPP is presently limited to low particle Reynolds number which will not be a general limitation of the method presented in this work. In concluding this literature review, there is much work being done to develop methods to estimate undisturbed quantities in a variety of scenarios (momentum/heat transfer, quasi-steady/unsteady sources, unbounded/wall-bounded flows etc.) and this paper is among those efforts to propose remedies to this computational challenge as well as elucidate deeper understanding of this problem.}

The remainder of the paper is broken into five parts. The first section demonstrates the relationship between the discretized fluid field equations and  discrete Green's functions, as well as how the discrete Green's function can be used to estimate the undisturbed fluid velocity. The next section discusses the procedure for finding discrete Green's functions. We apply this procedure to obtain the discrete Green's functions to the Stokes equations in a two-plane channel geometry. In the third section, we present a set of verification exercises where a particle settles under gravity moving parallel to a plane wall. We repeat this exercise for different wall-normal separations, particle Reynolds numbers, and methods to compute the undisturbed fluid velocity. In the fourth section, we consider a turbulent particle-laden channel flow to demonstrate that the manner of calculating the undisturbed fluid velocity may impact the turbulent statistics. Concluding remarks are given in the final section.
%\begin{center}
%\begin{sidewaystable}
\begin{table}
\centering
%\Rotatebox{90}{
\begin{tabular}{c c c c c c}
%\hline
$Method$ & $Coupling$ & $Reynolds \ \#$ &$Grid$& $Interpolation$ &$Projection$\\
% & \multicolumn{1}{c}{$Method$}
% & \multicolumn{2}{c}{$Reynolds number$}
% &
% & \multicolumn{3}{c}{$Grid$}
% &
% & \multicolumn{4}{c}{$Interpolation$} 
% 
% & \multicolumn{5}{c}{$Projection$} \\
%\hline
Gualtieri et al. 2015        & Mom & Stokesian & Anisotropic & General & Gaussian \\
Horwitz and Mani 2016        & Mom. & Stokesian & Isotropic                & General & Trilinear \\
Ireland and Desjardins 2017  & Mom. & Stokesian & Anisotropic & General & Gaussian \\
Horwitz et al. 2017          & Therm.  & Finite    & Isotropic                & General & Trilinear \\
Horwitz and Mani 2018        & Mom. & Finite    & Isotropic                & General & Trilinear \\
Esmaily and Horwitz 2018     & Mom. & Finite    & Anisotropic & General & General \\
Fukada et al. 2019             & Mom. & Finite    & Anisotropic & General & Gaussian \\
Balachandar et al. 2019            & Mom. & Finite    & Anisotropic & General & Gaussian \\
Poustis et al. 2019         & Mom. & Finite & Anisotropic & General & Gaussian \\
Liu et al. 2019            & Therm.  & Finite    & Anisotropic & General & Gaussian \\
%\hline
\end{tabular}
\caption{Different methods to compute the undisturbed velocity (momentum coupling) or temperature (thermal coupling). Interpolation refers to data transfer from fluid to particle while projection refers to transfer from particle to fluid. ``General''/ ``Anistropic'' refers to the fact that those particular methods may be suitable to a wider range of interpolation stencils and mesh configurations than specifically tested in those works.}
%}
\label{tab:err_bar}
\end{table}
%\end{sidewaystable}
%\end{center}
%\clearpage

\section{Origin of discrete Green's functions}
\label{sec:dgf_origin}
In this section, we demonstrate the connection between discretized partial differential equations and their associated discrete Green's function. We will focus our attention on the discrete Stokes equations. However, it is worth noting the method of discrete Green's functions applies to a much larger class of problems. \cite{Malgrange_1956} and \cite{Ehrenpreis_1954} proved that every linear partial differential equation with constant coefficients admits a Green's function. \cite{Zeilberger_2011} later showed every linear constant coefficient partial difference equation has a discrete Green's function. While Lewy's example \citep{Lewy_1957} demonstrates that not all non-constant coefficient linear partial different equations have a solution,  there still exists linear non-constant coefficient PDEs which have Green's functions, e.g. the spherically symmetric wave \citep{Haberman} and heat \citep{Hahn_Ozisik} equations. Therefore, while we focus on the Stokes equations in this section, the same manipulations can be readily performed for the thermal energy or Maxwell's equations e.g. written in Cartesian or other coordinate systems, as the application demands.

Let us consider now the steady Stokes equations, understood as the regular limit of the Navier-Stokes equations as the Reynolds number goes to zero, subject to a point-force density:

\begin{equation}
\label{Stokes_1}
\mu\nabla^2u_i-\frac{\partial p}{\partial x_i}= F_{i} 
\end{equation}

\begin{equation}
\label{Stokes_2}
\nabla^2 p = -\frac{\partial F_i}{\partial x_i}
\end{equation}

In the above $u_i$ and $p$ are respectively the fluid velocity and pressure, and $F_i$ is a force density applied at the particle center owing to two-way coupling. Let us consider a spatial discretization of the above system:

\begin{equation}
\label{Stokes_1_discrete}
\mu\frac{\delta^2u_i^n}{\delta x_j \delta x_j}-\frac{\delta p^n}{\delta x_i}= F_{i}^n 
\end{equation}

\begin{equation}
\label{Stokes_2_discrete}
\frac{\delta^2 p^n}{\delta x_j \delta x_j} = -\frac{\delta F_i^n}{\delta x_i}
\end{equation}

Here, the superscript $n$ denotes a field variable evaluated at gridpoint $n$ and $\frac{\delta}{\delta x_i}$ is the discrete differential operator associated to the continuous differential operator $\frac{\partial }{\partial x_i}$. By introducing the following discrete differential operators: $C^{kn} \equiv (\nabla^2)^{-1}$, $A_j()_j\equiv \frac{\partial }{\partial x_j}()_j$, and $B_{i} =\mathbf{\nabla}$, we can write the formal solution to Eqs. (\ref{Stokes_1_discrete}) and (\ref{Stokes_2_discrete}) as:

\begin{equation}
\label{discrete_pressure}
p^k = -C^{kn}A_jF_j^n\footnote{This discrete pressure solution is not unique since the uniform background pressure is arbitrary, however, the discrete pressure \textit{gradient} which is present in Eq. (\ref{Stokes_1_discrete}) is unique.}
\end{equation}

\begin{equation}
\label{discrete_velocity}
u_i^n = \mu^{-1}C^{nk}(\delta_{ij}^{kp}-B_iC^{kp}A_j)F_j^p
\end{equation}

By letting $G_{ij}^{np} = \mu^{-1}C^{nk}(\delta_{ij}^{kp}-B_iC^{kp}A_j)$, Eq. (\ref{discrete_velocity}) can be written in a simpler form:

\begin{equation}
\label{DGF_equation}
u_i^n = G_{ij}^{np}F_j^p
\end{equation}

In Eq. (\ref{DGF_equation}), $G_{ij}^{np}$ is called the discrete Green's function associated to the discrete Stokes equations. This discrete Green's function provides a mapping of a discrete point-force at grid point $p$ in direction $j$ to a discrete disturbance flow at grid point $n$ in direction $i$. Here, we have made no assumption regarding the imposed boundary conditions (homongeneous or wall-bounded e.g.), grid structure (uniform/non-uniform/unstructured), or discretization scheme ($2^{nd}$ vs. $4^{th}$ order finite differences, or spectral e.g.). In other words, there exists in general a different discrete Green's function for each discrete partial differential equation, for each set of boundary conditions, grid structure, and discretization scheme.

Our focus is not to explore sensitivity to each of these parameters but to elucidate how the discrete Green's function can be used to estimate undisturbed field quantities. In EL simulation, the force that a particle experiences $f_{i}$ is transferred from the Lagrangian particle to Eulerian discrete grid points via a projection operator, often via a Gaussian, Lagrange polynomial, or spline function. Dividing by the volume of the fluid element yields the discrete force density $F_i^n$. When this discrete force density is applied to the discrete fluid equations, the fluid velocity at each grid point will change from an undisturbed state $\tilde{u}_i^k$ to some new fluid velocity $u_i^k$, such that the new fluid velocity at gridpoint $k$ can be written in the form $u_i^k = \tilde{u}_i^k + u^{d,k}_i$, where $u^{d,k}_i$ is the disturbance velocity introduced by the particle at gridpoint $k$. If the disturbance velocity were known, then it would be possible to rearrange the previous expression to compute the undisturbed fluid velocity as $\tilde{u}_i^k = u_i^k - u^{d,k}_i$. The undisturbed fluid velocity at the particle location could then be calculated by interpolation of the undisturbed fluid velocity from the Eulerian grid points to the particle location. Let us introduce the operators $W^k$ and $w^n$ as the weights associated with interpolation and projection, respectively. If we assume that the discrete Green's function previously introduced provides a good approximation between the discrete force density $F_i^n$ and discrete disturbance flow $u^{d,k}_i$, the undisturbed fluid velocity at the particle location can be estimated as:

\begin{equation}
\label{estimate_undisturbed}
\tilde{u}_i \approx W^k\big\{u^k_i-G_{ij}^{kn} w^n f_j\big\}
\end{equation}

Interestingly, the undisturbed fluid velocity can be accurately estimated regardless of the choice of interpolation/projection stencil as previously alluded to in \cite{Esmaily_Horwitz_2018}. While Gaussian schemes may offer certain advantages to stability or smoothing of flow-field near the particle \citep{Poustis_etal_2019}, it seems low-memory access interpolation/projection may still be advantageous since these EL data-transfer methods still allow accurate computation of the undisturbed fluid velocity and hence the energetics of the system will be consistent with the chosen drag law \citep{Horwitz-2019,Mehrabadi_2018}. From the standpoint of storage, it may also be favorable to use low-memory access projection/interpolation stencils to reduce the cost of storing the discrete Green's function. For even a modest problem size of say $10^{6}$ gridpoints, the whole discrete Green's function would involve $O(10^{12})$ elements for a general problem. However, if we consider only the points which will actually be accessed for computing the undisturbed fluid velocity, the $n$ points associated with $W^n$ and $w^n$, the result can be enormous computational savings. We will give a more precise estimate of the storage and computational cost associated with discrete Green's functions in the next section.
%{\color{red} figure out notation force density vs. force factor of volume is missing}

\section{Finding discrete Green's functions}
\label{sec:find_dgf}
For a general problem, the discrete Green's function for the Stokes equations involves the mapping of $3$ force components at $N$ gridpoints, % in the simulation,
 to $3$ velocity components at $N$ gridpoints, so that the total size of the DGF is $9N^{2}$. In addition to the large memory requirements to store this object, the construction of the DGF would require solving the Stokes equations successively, in each case with a point-force placed at a different gridpoint in a different coordinate direction. Therefore, direct computation of the DGF is infeasible for most problems. An alternative approach for general problems which relies on approximating DGF's locally is discussed in \cite{Horwitz_Thesis_2018}.

\begin{figure}
  \centering
  \subfigure[]{\includegraphics[trim={0.5cm 4.5cm 0.5cm 4.0cm},clip, height=0.32\textwidth,page=1]{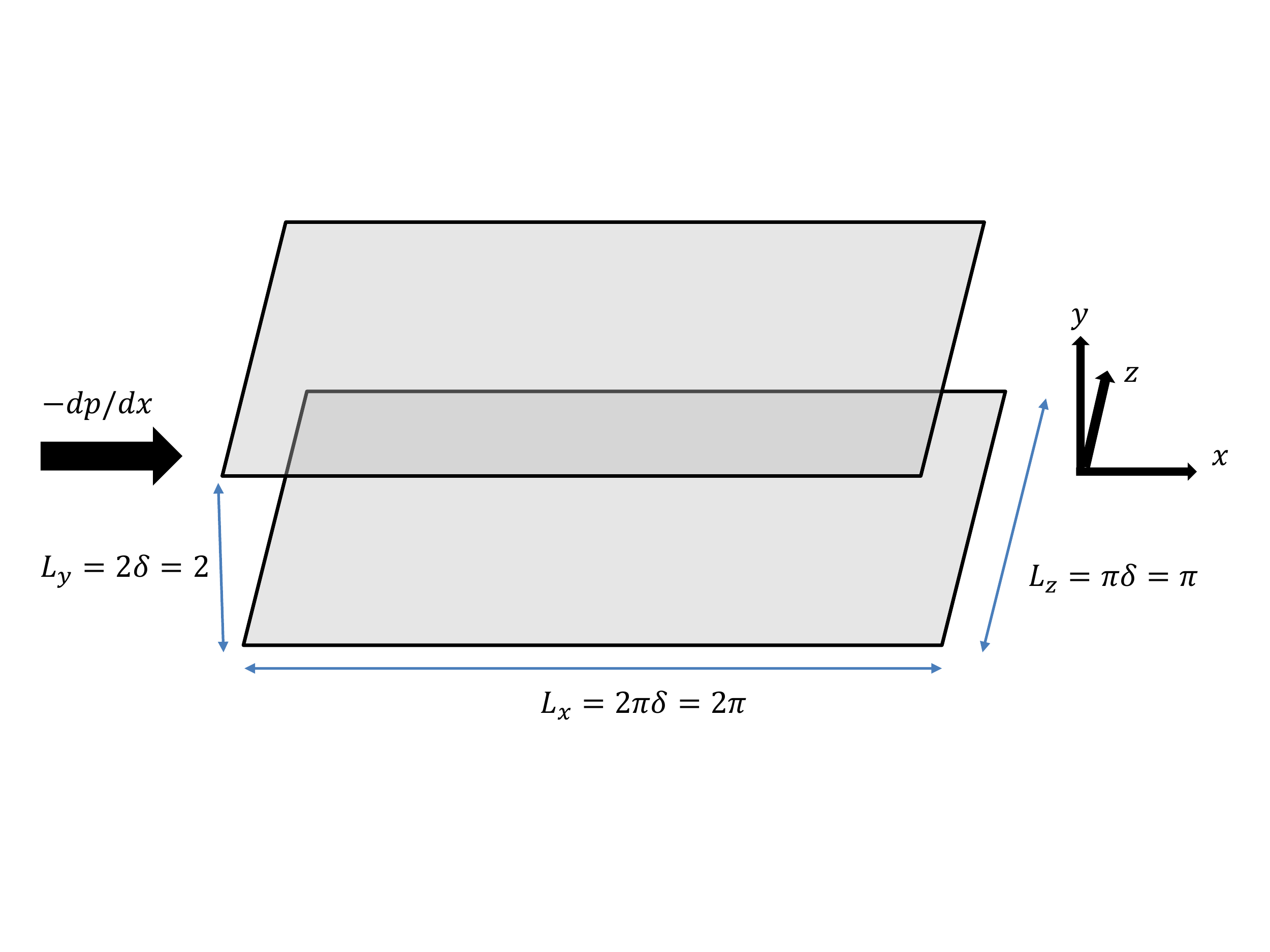} \label{fig:kf_100}}  
  \centering
  \subfigure[]{\includegraphics[trim={0.5cm 0.0cm 1.25cm 0.0cm},clip, height=0.42\textwidth,page=2]{figures/Plane_Channel_Figures.pdf} \label{fig:ef_100}}  \caption{(a) Simulation domain and (b) depiction of the locations where point-forces will be placed in order to calculate the discrete Green's functions.}
  \label{fig:geometry}
\end{figure}

\begin{figure}
  \centering
  \subfigure{\includegraphics[trim={4.7cm 4.0cm 4.5cm 3.0cm},clip, height=0.55\textwidth,page=4]{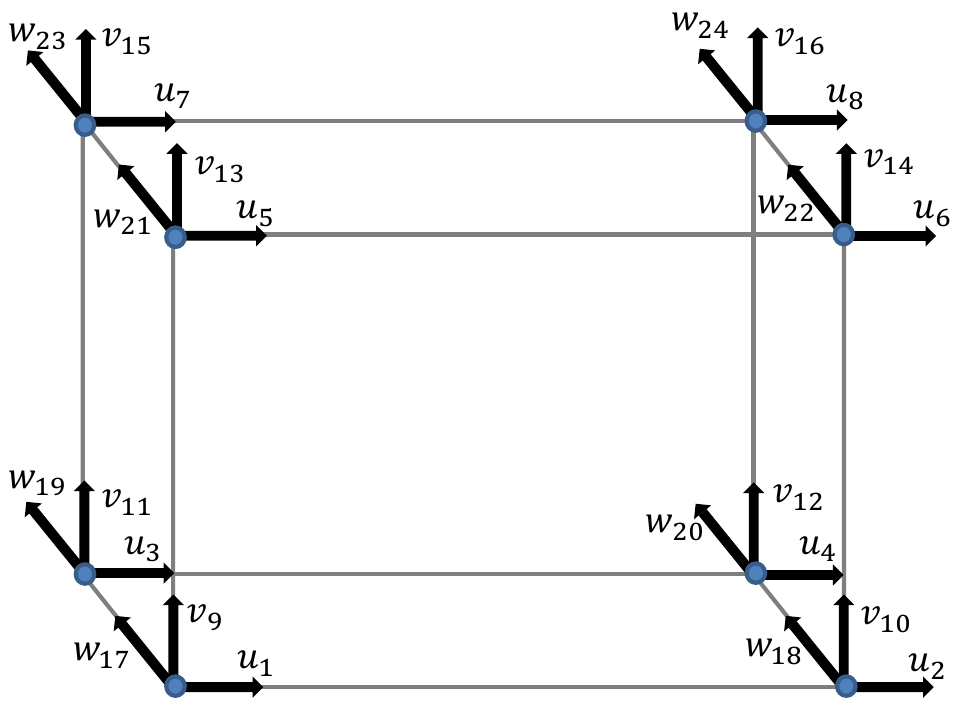} \label{fig:24_gridpoints_1}}  
\caption{Gridpoints involved in trilinear interpolation and projection {\color{black}for a collocated arrangement}. For an arbitrary location of a particle in the domain, the surrounding gridpoints can be mapped to the notation used here. Streamwise components are denoted with $u$, spanwise with $w$, and wall-normal with $v$.}
\label{fig:24_gridpoints}  
\end{figure}

In this work, we will consider the computation of the discrete Green's function for a two-plane channel flow geometry, a domain with considerable interest in particle-laden flow problems \citep{Kulick_1994,Sardina_2012,Brandt_2019}. In addition, the two-plane channel geometry has two homogeneous directions which may be exploited to dramatically reduce the cost associated with computation and storage of the DGF. A diagram of the flow domain is shown in Figure~\ref{fig:geometry} (a). The boundary conditions are periodic in the streamwise (x) and spanwise (z) direction, while the velocity obeys no-slip and no-penetration at the top and bottom walls. Owing to the homogeneous directions (x,z), the discrete Green's function will be invariant with respect to translation in those directions. Hence we may choose a fixed (x,z) line to consider the problem of finding the discrete Green's function for different wall-normal particle positions (Figure~\ref{fig:geometry}(b)). Owing to symmetry about the centerplane of the channel, we need only consider one half of the channel. Therefore, to compute the whole Discrete Green's function will require running $3N_{y}/2$ simulations of the Stokes equations in the two-plane channel geometry. By assuming the projection and interpolation operators are trilinear stencils, we will ultimately construct an object of size $24\times24\times N_{y}/2$. In a trilinear stencil, eight {\color{black}degrees of freedom} %grid points 
are involved for each of three velocity directions. In other words, for a given wall normal distance a general force which may have components in the x, y, and z directions will involve $24$ {\color{black}degrees of freedom}, %grid points, 
and will be mapped, at a given wall-normal distance, through a tensor involving $24\times 24 = 576$ components, to $24$ velocity {\color{black}components}, %gridpoints, 
eight for each of three directions. Only these $24\times 24$ elements must be considered for each wall-normal location because the choice of trilinear stencil to be incorporated in the interpolation/projection stencils ensures there will be no contribution to the self-disturbed fluid velocity at a given particle's location owing to distant gridpoints, or those outside the set of $24$. {\color{black}See Figure \ref{fig:24_gridpoints} which shows a collocated grid arrangement to make the surrounding discussion more intuitive. The staggered grid implemented in the present work requires additional considerations which are discussed in Appendix B.}

%{\color{red} (a)figure two-plane channel geometry (b) depiction of 3Ny/2 simulations in wall normal direction}

\begin{figure}
  \centering
  \subfigure[]{\includegraphics[trim={3.4cm 8.0cm 4.7cm 7.5cm},clip, height=0.44\textwidth]{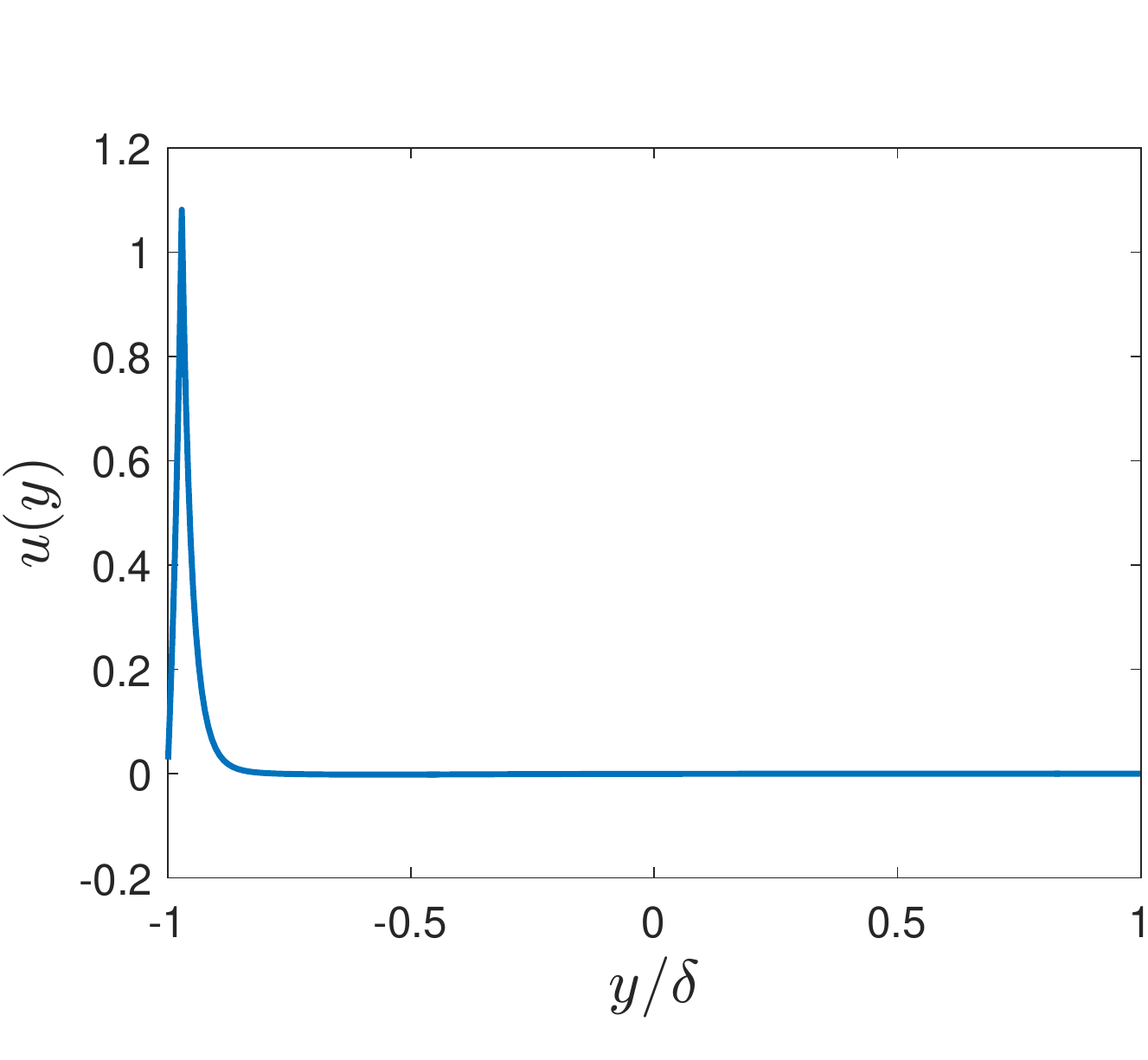} \label{fig:kf_100}} 
  \put(-145,74){\includegraphics[trim={4.4cm 8.8cm 4.3cm 9.0cm},clip, width=45mm]{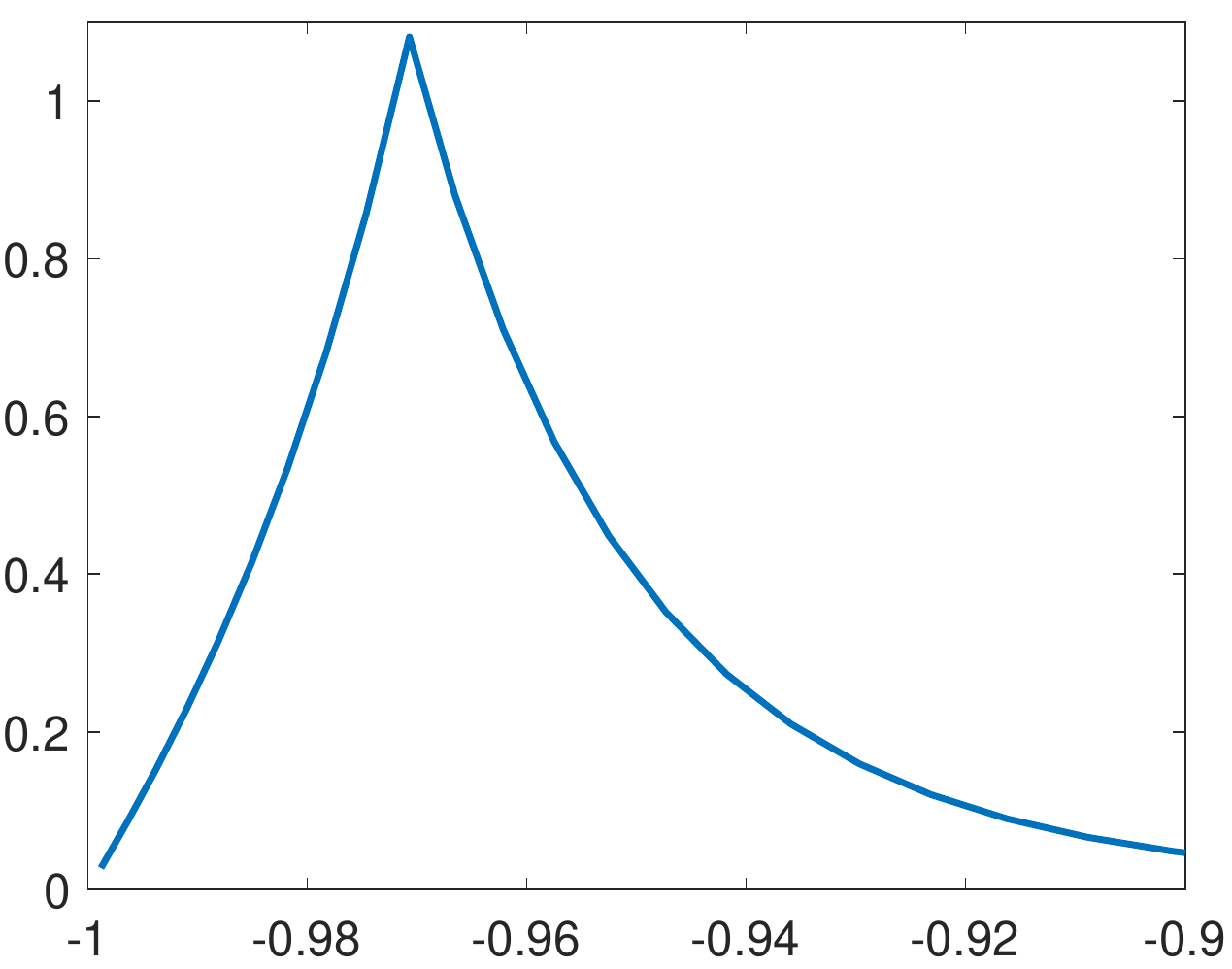}} 
  \centering
  \subfigure[]{\includegraphics[trim={3.4cm 8.0cm 4.7cm 7.5cm},clip, height=0.44\textwidth]{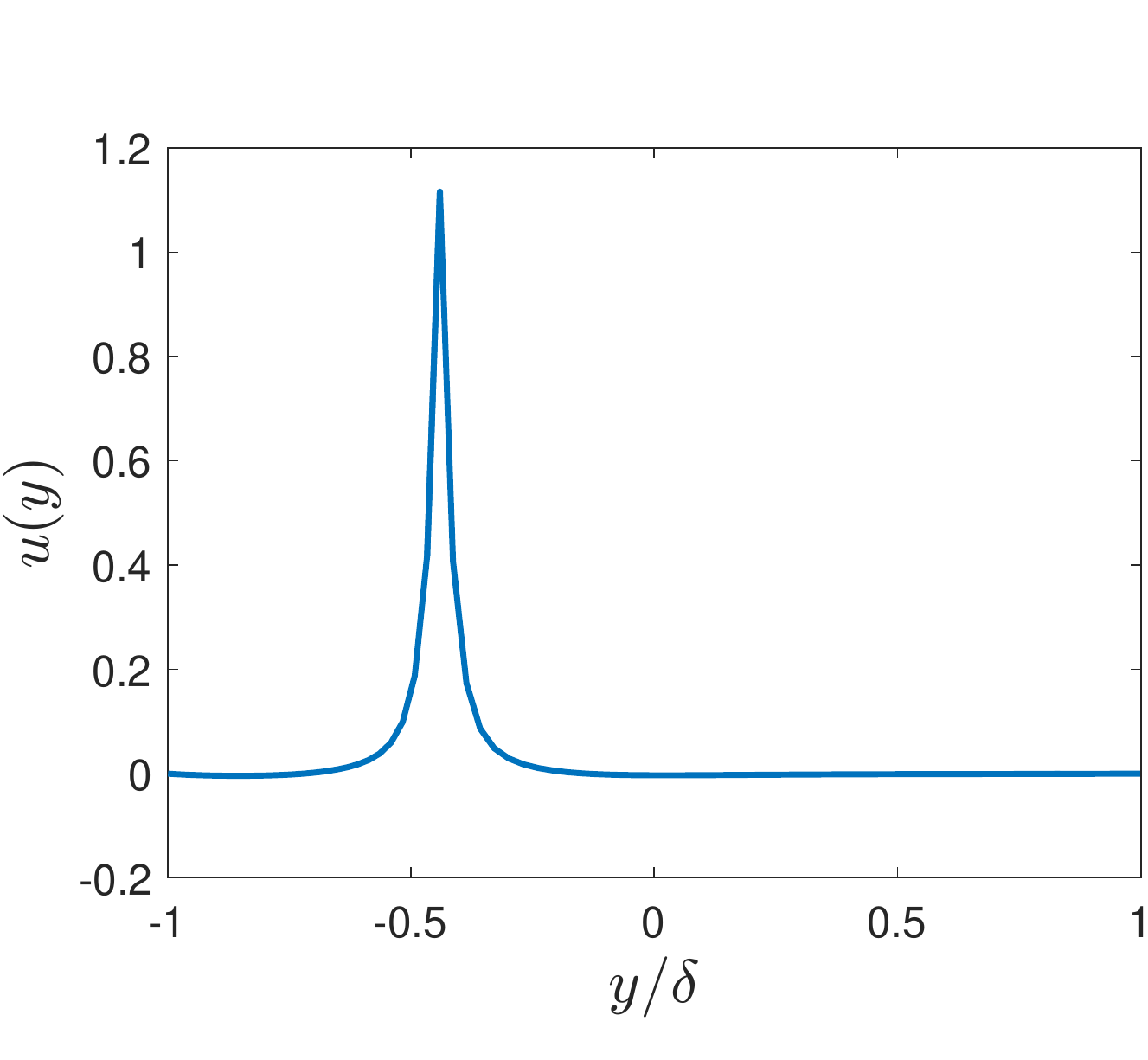} \label{fig:ef_100}}
  \put(-119,79){\includegraphics[trim={4.4cm 8.8cm 4.3cm 9.0cm},clip, width=40mm]{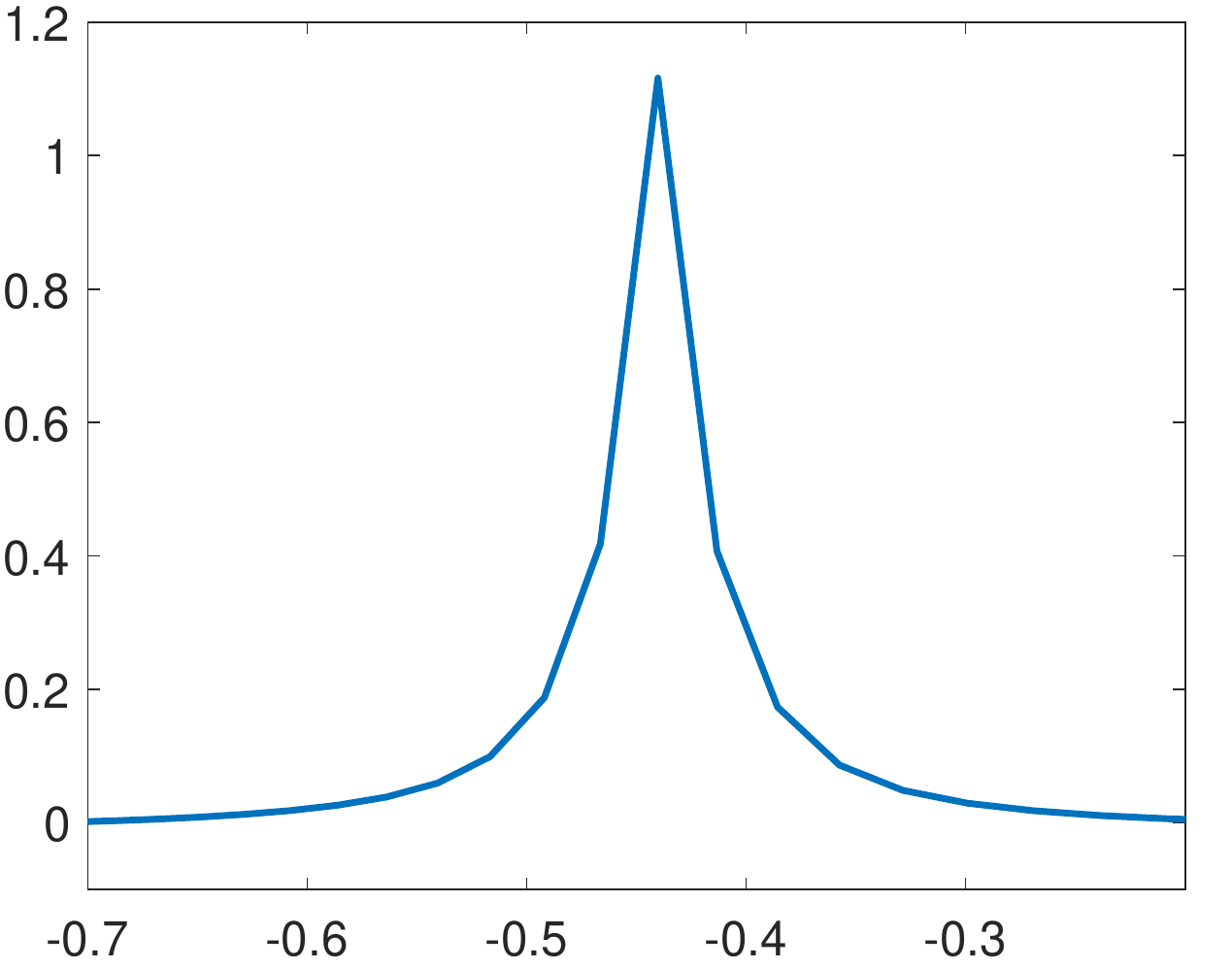}} 
    \caption{Profiles of stream-wise velocity disturbances generated by stream-wise point-force placed in the (a) near wall, and (b) away from wall regions. Insets show near-field response. Notice the strong assymetry in the stream-wise response in the near-wall disturbance compared with the away-from wall disturbance. }
  \label{fig:velocity_spike}
\end{figure}

To calculate the DGFs, the Navier-Stokes equations are solved at low Reynolds number and integrated for several viscous times until the solution becomes steady. This process is repeated $3N_y/2$ times; in each simulation, a point-force of magnitude unity is placed at different grid points in the domain, one for each coordinate direction, and $N_y/2$ to cover all wall-normal locations. (This problem is symmetric about the channel center-plane so we need only consider half of the wall-normal grid points.) The magnitude of the kinematic viscosity $\nu$ and density $\rho$ is chosen such that $f/2\pi\rho\nu^2 \ll 1$, so that this discrete problem represents the low-Reynolds number limit of the Landau-Squire jet \citep{batchelor-1967}. The mean stream-wise pressure gradient, $-dp/dx$ is set to zero, so that the resulting flow is only owing to the presence of the discrete point-force. The streamwise and spanwise lengths are large enough to ensure minimal effects from the periodic boundary conditions (see the disturbance length scale in Figure~\ref{fig:velocity_spike}). Time-advancement is accomplished with 2nd order Runge-Kutta. The grid arrangement is a staggered configuration (Figures~\ref{fig:geometry}(b)). %and \ref{fig:24_gridpoints}). 
The grid spacing is uniform in the stream-wise and span-wise direction with 96 gridpoints in each of those directions. The wall-normal direction uses 128 gridpoints in a non-uniform arrangement given by a hyperbolic tangent profile which clusters more gridpoints near the wall. The pressure equation is solved using multi-grid acceleration. All spatial derivatives are discretized with 2nd order finite differences. More details on the implementation can be found in \cite{Esmaily_Jofre_2018}.

\begin{figure}
  \centering
  \subfigure[]{\includegraphics[trim={0.9cm 6.7cm 2.6cm 7.5cm},clip, height=0.36\textwidth]{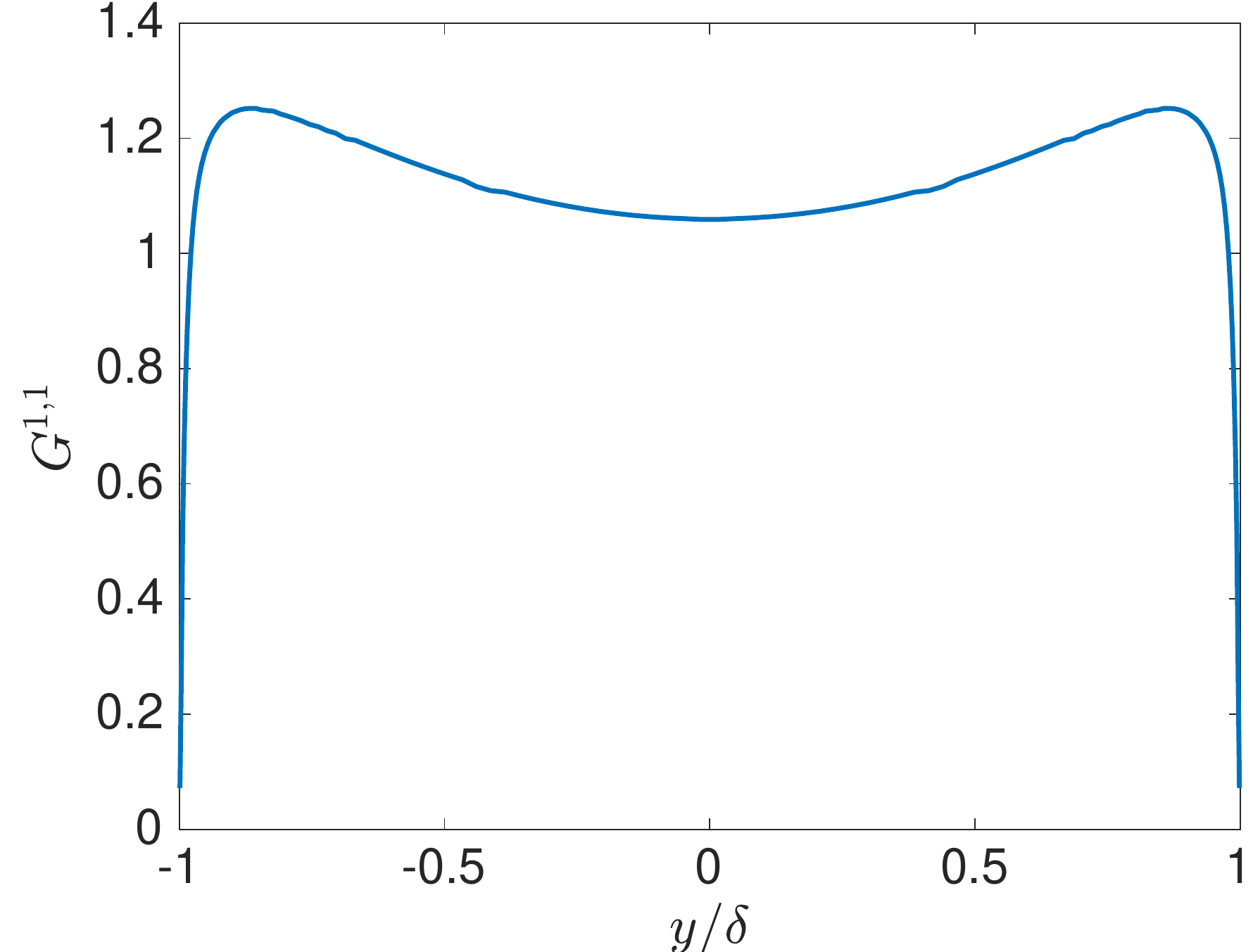} \label{fig:kf_100}}  
  \centering
  \subfigure[]{\includegraphics[trim={0.9cm 6.7cm 2.6cm 7.5cm},clip, height=0.36\textwidth]{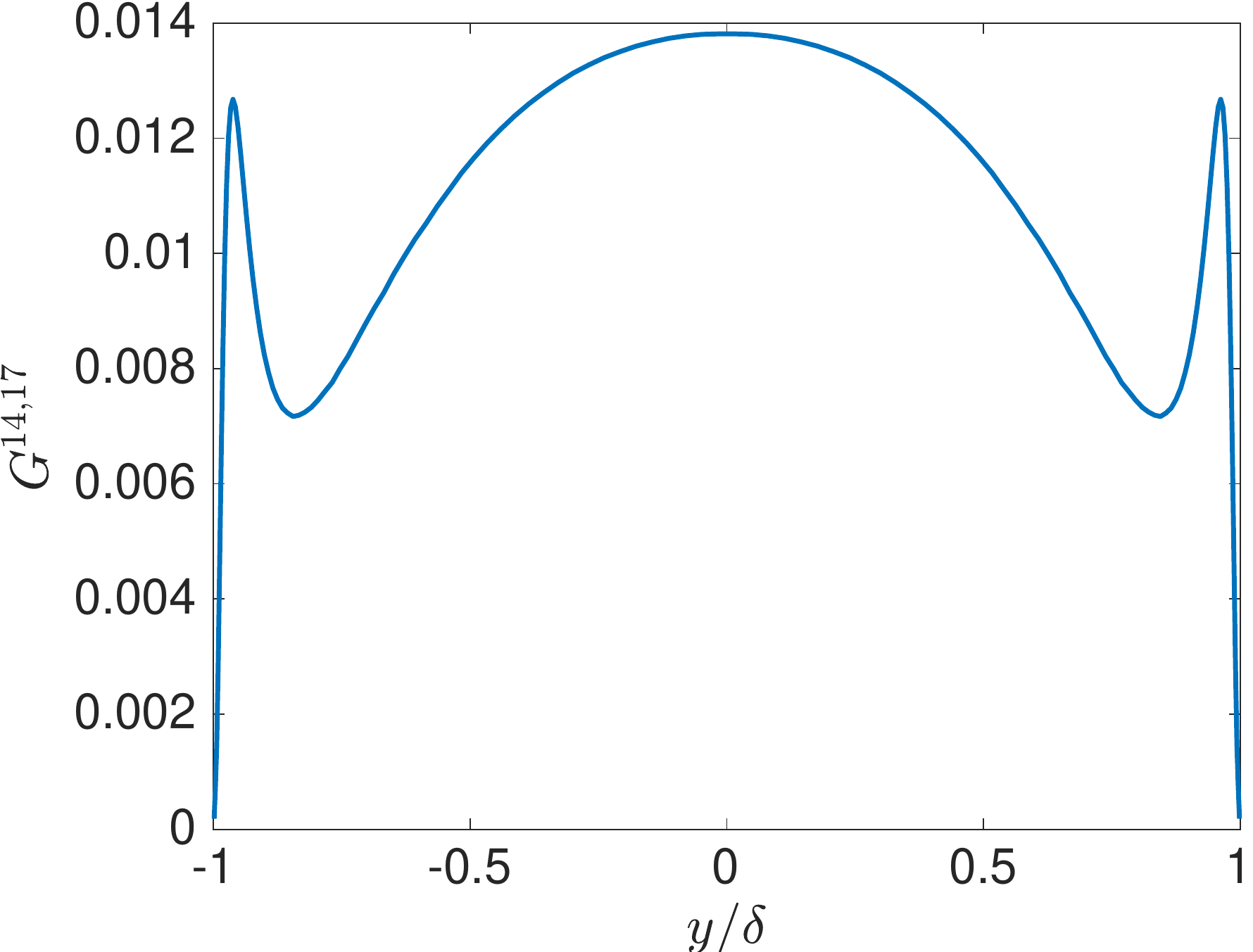} \label{fig:ef_100}}  \caption{Wall-normal profiles of the locus of selected DGF elements, (a) $G^{1,1}$, (b) $G^{14,17}$. {\color{black}The notation for these elements is provided in Appendix B in Figures~\ref{fig:24_staggered}.}}
  \label{fig:locus}
\end{figure}

\begin{figure}
  \centering
  \subfigure[]{\includegraphics[trim={2.3cm 7.4cm 2.3cm 7.5cm},clip, height=0.37\textwidth]{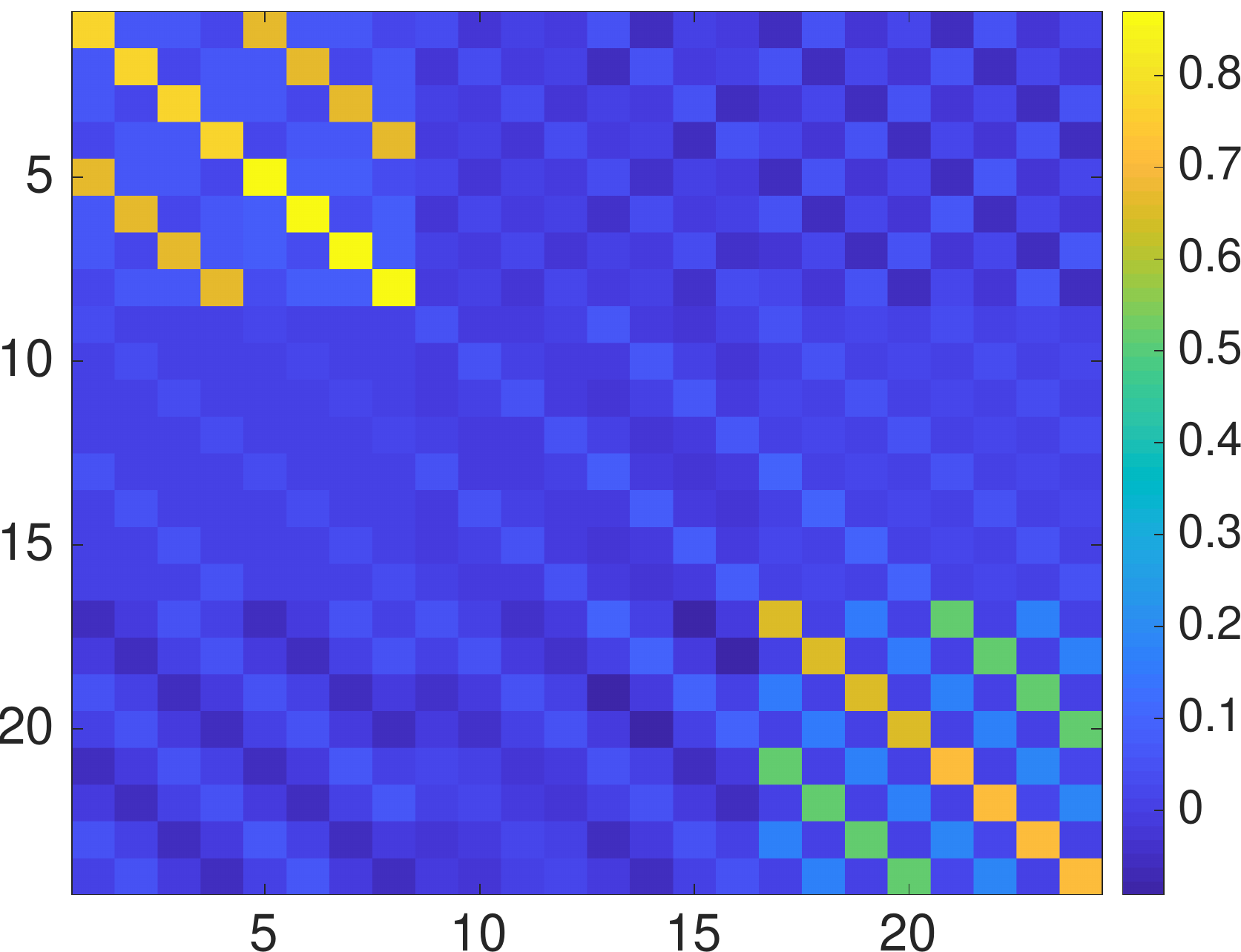} \label{fig:contour_wall}}  
  \centering
  \subfigure[]{\includegraphics[trim={2.3cm 7.4cm 2.3cm 7.5cm},clip, height=0.37\textwidth]{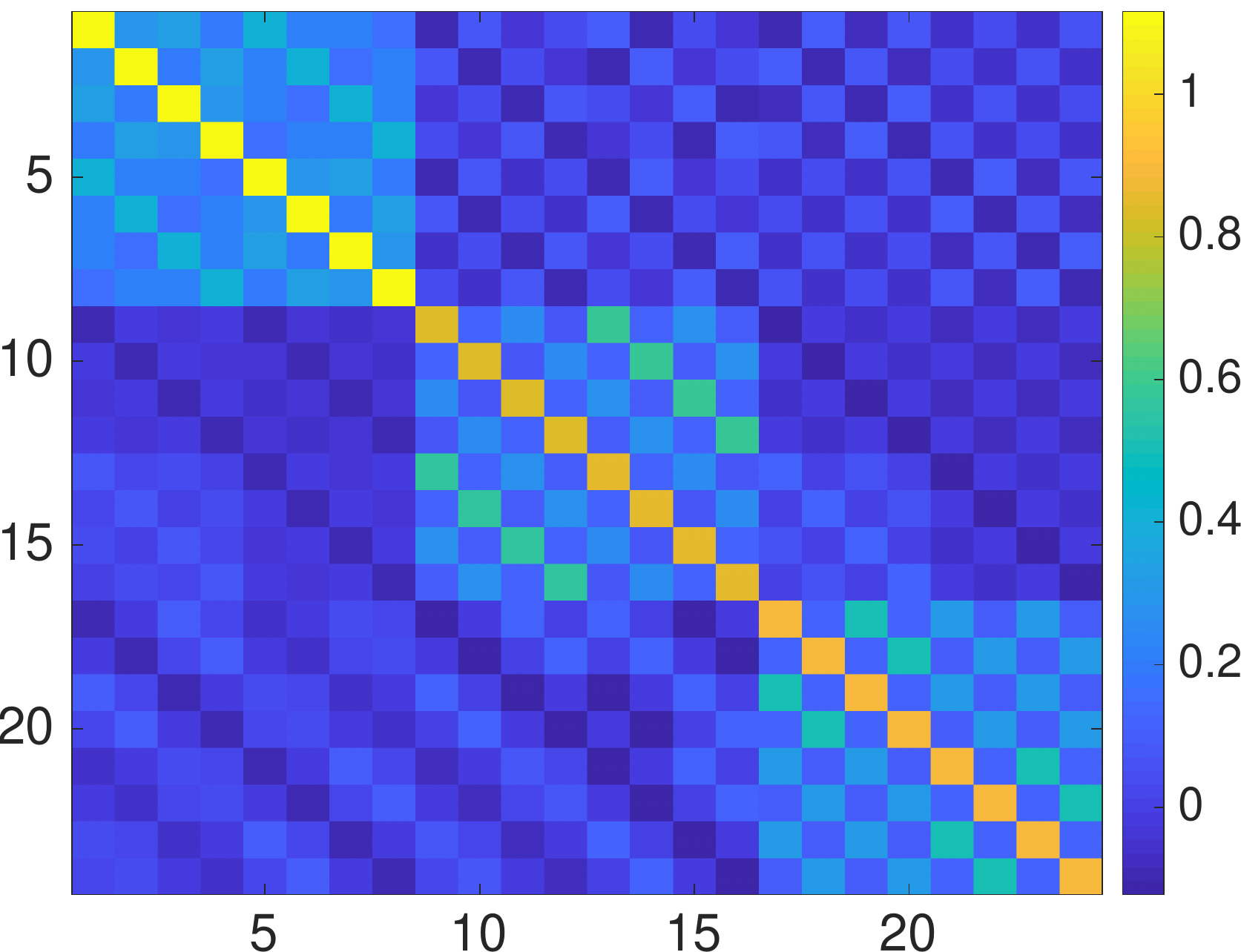} \label{fig:contour_center}}  \caption{Elements of DGF at different wall-normal locations, (a) near-wall, (b) near center. {\color{black}These elements are extracted from a staggered configuration which is the basis of the flow solver used in this work. The notation for these elements is provided in Appendix B in Figures~\ref{fig:24_staggered}.}}
  \label{fig:contours}
\end{figure}

To determine the discrete Green's functions from the calculated solution, we can isolate each of the elements of the DGF for each wall-normal distance. For example, the velocity disturbance %in the 
for stream-wise %direction 
{\color{black} component} %at gridpoint 
$1$ {\color{black}owing to} %for 
a point-force placed at {\color{black}component} %gridpoint
 $1$, can be obtained as $G^{1,1}=u_{1}/F_{1} = u_1$ (see Figure~\ref{fig:24_gridpoints}). Similarly, for a point-force applied at %gridpoint 
 $9$, $G^{9,9}$ can be calculated as $G^{9,9}=v_{9}/F_{9}=v_{9}$. Certain off diagonal {\color{black}elements of the DGF} %components 
 can be directly measured using this strategy. For example, when the force is applied to %grid-point 
 $1$, the velocity calculated at %gridpoint 
 $5$ corresponds to $G^{5,1}$. However, not all elements of the DGF can be directly calculated from the $3N_{y}/2$ simulations. For these remaining  components of the DGF, it is necessary to exploit symmetries. For example, in practice we only put a force at %gridpoint 
 $1$ and not %gridpoint 
 $2$. However, had we put a force on $2$, it can be observed that $G^{2,2}=G^{1,1}$. %Another type of symmetry which exploits 
{\color{black}Stokesian symmetry can also be exploited, for example} the response at %gridpoint 
$1$ due to a force applied to %gridpoint 
$4$ is the same as the response at $4$ due to a force at %gridpoint 
$1$, viz $G^{1,4}=G^{4,1}$. By exploiting all such symmetries, we can fill out the DGF for a given wall-normal location. By then combining the 576 elements for each wall normal location, we develop the complete DGF object for the channel flow. In Figure \ref{fig:locus}, we plot the locus of $G^{1,1}$ and $G^{14,17}$ as a function of wall-normal distance to gain an understanding of what the DGF object looks like. The elements of the DGF are symmetric about the channel center-plane. In addition, we can see the magnitude of $G^{1,1}$ is much greater than the magnitude of $G^{14,17}$ over the whole of the channel. It is not surprising that the greatest response to the point-force is typically at the location of application of the point-force and in the same direction. There is an exception, however, with respect to the wall-normal diagonal components of the DGF. This is reflected in Figure~\ref{fig:contours}. Here we have plotted all components of the DGF at two wall-normal locations, one close to the wall in~\ref{fig:contours} (a), and one close to the center of the channel (Figure~\ref{fig:contours}(b)). Near the wall, when a force is applied in the wall-normal direction, there is minimal response. The no penetration condition demands the induced wall-normal response to scale as $y^{2}$ near the wall \citep{pope-2000}. In contrast, the no-slip condition demands that the stream-wise and span-wise velocity disturbances vary linearly near the wall. Hence, force components in the stream-wise and span-wise directions generate relatively large responses in those directions compared with the wall-normal direction. In other words, the velocity disturbance induced by an arbitrary point-force near the wall is almost planar. In contrast, near the center of the channel, the wall-damping effects are diminished, and the velocity disturbance owing to an arbitrary point-force is a three-dimensional, but diagonally dominant, response (Figure~\ref{fig:contours}(b)). 

\begin{figure}
  \centering
  \subfigure{\includegraphics[trim={0.8cm 6.7cm 2.7cm 7.3cm},clip, height=0.48\textwidth]{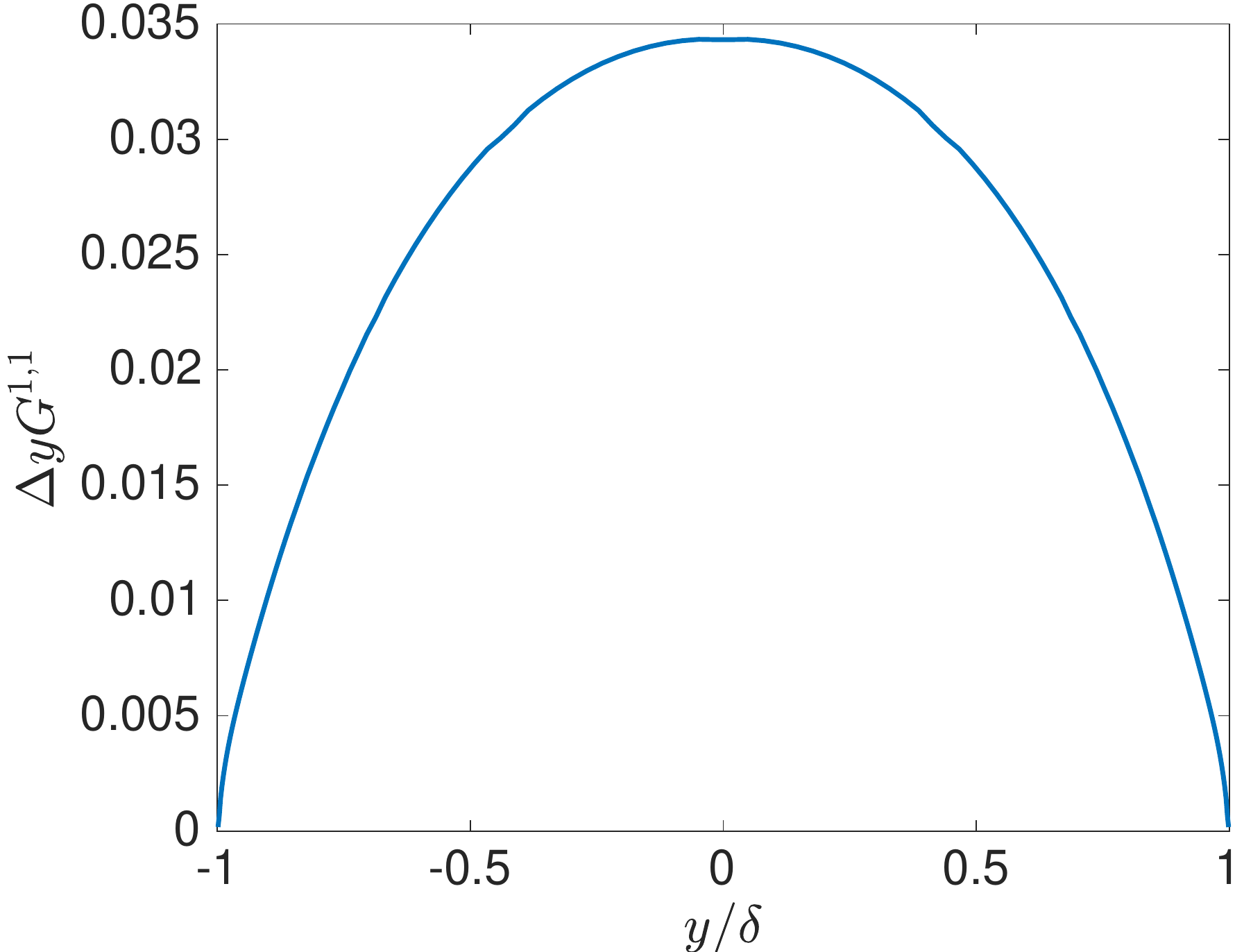} \label{fig:kf_100}}  
  \caption{$G^{1,1}$ profile compensated by local wall-normal grid-spacing. {\color{black}The notation for these elements is provided in Appendix B in Figures~\ref{fig:24_staggered}.}}
  \label{fig:normalized_locus}
\end{figure}

The shapes of the DGF profiles (Figure~\ref{fig:locus}), particularly that of $G^{1,1}$, are not intuitive. It would be expected for the response to be weakest near the wall and monotonically transition to the largest response being near the center. {\color{black}This non-monotonic variation can be seen as an artifact (but purposeful choice) of explicitly building in the effect of an arbitrary grid spacing, which is non-uniform in the wall-normal direction for this example. When the Lagrangian point-force is translated to a discrete force-density (see Eq. (\ref{estimate_undisturbed}), it is divided by the local elementary fluid volume $V = \Delta x \Delta dy \Delta dz$. For a given force, this factor of volume can be absorbed into the DGF. To elucidate this observation, if we re-normalize $G^{1,1}$ by $1/\Delta y$, it is clear that the effect of the grid-spacing is removed and the profile exhibits the expected behavior (Figure \ref{fig:normalized_locus}).}

While the procedure described in this section for obtaining and storing the DGFs may appear costly, it still may be practical for simple geometries like the two-plane channel. The total number of DGF elements $24\times 24\times 128 = 73, 728$ is small in comparison to the total number of gridpoints for the present simulation $96\times 96 \times 128 = 1, 179, 648$, so that there is no significant storage overhead involved in incorporating DGFs into a particle-laden turbulent channel simulation which would otherwise run without them (see Section~\ref{sec:turbulent_channel}). In addition, while running $3N_{y}/2$ simulations may seem costly in comparison to running one particle-laden channel simulation without DGF, obtaining the DGF's only requires running simulations for order of diffusion times while running a turbulent channel simulation requires considerably longer calculations, especially with particles whose steady wall-normal concentration profiles can be slow to develop \citep{Sardina_2012}. Implicit time-stepping can be used to obtain steady-state DGFs in reduced computational time. Smaller stream-wise and span-wise domains can also be exploited as well (these length-scales need not correspond to the application length-scale, they need just be an order of magnitude larger than the projection-operator stencil bandwidth. See the response curves in Figure \ref{fig:velocity_spike}). All of these considerations together mean that computing and storing the DGF's for the two-plane channel flow, then subsequently incorporating these DGF's to more accurately compute particle motion in a turbulent channel calculation, can be done with the same order of magnitude in simulation expense (simulation size $\times$ of time steps) as running a turbulent channel flow with particle motions less accurately computed, without the aide of DGF's \citep{Horwitz_Thesis_2018}. For general domains and grid structures, a careful analysis should be undertaken in each case to assess how the cost of obtaining/storing the DGF for that problem is balanced by the enhanced accuracy that would come from incorporating the DGF's into the application simulation.
%{\color{red}Find a place to add: the memory access requirements of the DGF, e.g. 24x24 is comparable to the 8th order Lagrange stencil which has been used as an interpolation stencil particle-laden simulation \citep{Ray_Collins_2011}.}
%{\color{red} figure streamwise response vs wall-normal coordinate (a)near channel center and (b)near wall (showing spike in velocity profile near location of point-force.)}

%{\color{red} locus of streamwise response G11 vs wall normal coordinate (a) un-normalized (b) normalized with V/mu}

%{\color{red} (a) 24x24 G matrix near wall and (b) near channel center}

%\clearpage
\section{Verification}
\label{sec:ppdns_method}

Having obtained the DGF for a two-plane channel geometry in the previous section, it is worth assessing whether this object can be used to estimate the undisturbed fluid velocity for a moving particle. This section is concerned with verification, namely, if we prescribe a force-model for a particle, do we get the motion of the particle consistent with that choice of force model. In this section, we consider the motion of a particle moving parallel to a plane wall, for different wall-normal separations. We prescribe a constant body force felt by the particle in a direction parallel to the wall, but not aligned with either the streamwise or span-wise directions. We consider a force-model of the form:

\begin{equation}
\label{F_general_form}
f_{i}=3\pi d_{p}\mu K(l/r)(\tilde{u}_i-v_i)(1+0.15Re_{p}^{0.687})
\end{equation}

Eq. (\ref{F_general_form}) has been adopted to account for two considerations. The
$(1+0.15Re_{p}^{0.687})$ term is a finite particle Reynolds number extension to Stokes drag, while $K(l/r)$ is a wall-correction factor which accounts for the variation in the drag a particle experiences near a wall compared with the drag it would experience in a uniform un-bounded flow. Here, $l/r$ is the ratio of wall-normal distance to particle radius. To the authors' knowledge, outside of low particle Reynolds number, where Eq. (\ref{F_general_form}) reduces to $f_{i}=3\pi d_{p}\mu K(l/r)(\tilde{u}_i-v_i)$, there is no strict justification to adopt a force model of this form. Instead, we have chosen to adopt this form to highlight different physical interactions particles may experience in the vicinity of the wall. In the context of verification, this form is arbitrary. However, to appeal to the origin of these two extensions to Stokes drag, we only consider a non-unity value of $K$ in the low particle Reynolds number verification study.

In addition to considering particle Reynolds number, wall-correction, and wall-normal distance as variables, we also compare the DGF method developed in this work with two existing methods to estimate the undisturbed fluid velocity. The first method simply uses trilinear interpolation to estimate the undisturbed fluid velocity at the location of the particle. Effectively this means, using Eq. (\ref{estimate_undisturbed}) with just the first term. Trilinear was chosen as an appropriate interpolation scheme because it has been found to be more accurate than higher-order interpolation schemes in the context of methods which measure the disturbed fluid velocity as an estimate for the undisturbed fluid velocity \citep{Horwitz_2016,horwitz-2016b}. The other method is a recent correction scheme developed by \cite{Esmaily_Horwitz_2018} which has been verified on anisotropic grids in \textit{unbounded} flows. For all cases, a particle is released from rest at a prescribed wall-normal location. The particle velocity reaches a steady state when the drag and imposed body force on the particle are balanced. We then compare the particle time-averaged velocity to the terminal velocity which would be consistent with the prescribed constant body force balancing the drag, with the analytical undisturbed fluid velocity, $\tilde{u}_i = 0$. In other words, the reference settling velocity $u_s$ satisfies $3\pi d_{p}\mu Ku_s\big(1+0.15(u_s d_p/\nu)^{0.687}\big) = F_{body}$. The particle diameter chosen leads to non-dimensional size ratios of $d_p/\Delta x \approx 0.2$, $d_p/\Delta y_{min} \approx 5.5$, $d_p/\Delta y_{max}\approx 0.4$, $d_p/\Delta z \approx 0.4$, $d_p/\delta = 0.013$, making the present verification realistic in terms of typical applications though challenging from the standpoint that an undisturbed fluid velocity correction is needed for this configuration.

\subsection{low $Re_{p}$ verification}
{
\label{sec:lowRepverifcation}
In this section we consider the verification problem described previously for particle Reynolds $Re_{p} = 0.05$. Plotted in Figure~\ref{lowRepverification} is the percent error for particle settling velocity (velocity parallel to the wall) compared with the reference velocity for a given drag force model. For three of the methods: ``Trilinear Interpolation'', ``Esmaily~\&~Horwitz,~JCP~2018,'' and ``DGF'', we use a less realistic wall model with $K(l/r) = 1$. The ``DGF + Brenner'' case incorporates discrete Green's function combined with the \cite{Brenner_1962} correlation for drag coefficient, $K(l/r) = \big(1-(9/16)(r/l)\big)^{-1}$ which agrees well with the exact solution of \cite{Oneil_1964} even at distances $r/l = O(1)$. As expected, the trilinear interpolation method, which makes no distinction between the undisturbed and disturbed velocity, shows the highest error in particle settling velocity. The method developed by \cite{Esmaily_Horwitz_2018} is very accurate at predicting the undisturbed fluid velocity and therefore particle settling velocity throughout most of the channel. However, while this method has been verified for highly skewed grid cells which are characteristic of the cells near the wall in this setup, the \cite{Esmaily_Horwitz_2018} method was not developed to explicitly account for the velocity-damping which takes place in near-wall proximity. %To estimate where this increase error occurs, we note the final section of this paper (\ref{sec:turbulent_channel}), which concerns turbulent channel flow at $Re_{\tau}=180$, uses the same particle sizes, which in wall units corresponds to a non-dimensional particle size of $d_{p}^{+} \approx 2.3$. So the increase in error observed for the \cite{Esmaily_Horwitz_2018} method roughly corresponds to the viscous sublayer and part of the buffer layer, had this verification problem considered turbulent flow. 
On the other hand, the DGF approach predicts very small error at all wall-normal separations, including the closest case considered with $l/r \approx 3$. The ``DGF + Brenner'' case uses DGFs to predict the undisturbed fluid velocity as before, with a wall-correction factor $K(l/r)$ calculated using Brenner's formula. For reference, at the closest wall-normal separation considered, $K(l/r)\approx 1.22$. Despite this enhancement in drag, there is negligible change in the results. The DGF approach is able to accurately predict the undisturbed fluid velocity for this modified force-model which demonstrates that the form of the force model is not important, but rather the magnitude of the force satisfies the condition $f/2\pi \rho  \nu^{2}\ll 1$ consistent with the use of discrete Green's functions of the discrete Stokes equation \citep{batchelor-1967,Balachandar_2019a}. In other words, we have verified that the DGF model can accurately predict the undisturbed fluid velocity of a moving particle at low Reynolds number in wall-bounded settings, regardless of the form of force model, which may change based on various physical considerations. We speculate that Gaussian based schemes which estimate the undisturbed fluid velocity assuming unbounded settings \citep{gualtieri_etal_2015,ireland_desjardins_2017,Balachandar_2019a} would perform qualitatively like the \cite{Esmaily_Horwitz_2018} method, while the scheme developed by \cite{Battista_2019} would likely exhibit qualitatively similar performance to the DGF results presented here.
%{\color{red} figure: low Re verification}
\begin{figure}
  \centering
  \subfigure{\includegraphics[trim={3.5cm 7.6cm 3.5cm 8.2cm},clip, height=0.58\textwidth]{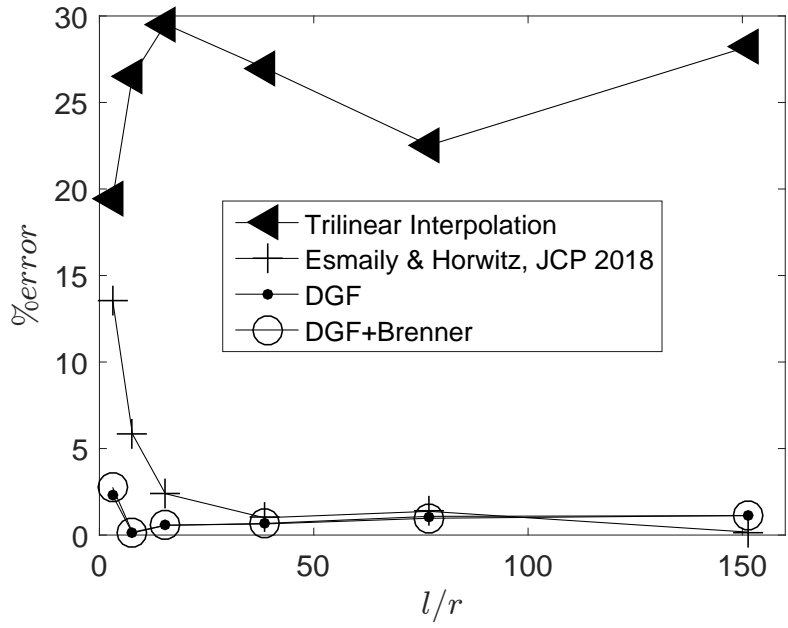} \label{fig:kf_100}}  
\caption{Percent error in settling velocity predicted by different schemes of a particle moving parallel to a plane wall for different wall-normal separations, $Re_{p} = 0.05$.}
\label{lowRepverification}
\end{figure}
}
%\clearpage
\subsection{finite $Re_{p}$ verification}
{
We next consider the settling verification problem described previously for particle Reynolds $Re_{p} = 1$ (Figure \ref{finiteRepverification}). All cases in this section use $K(l/r) = 1$. As with the low particle Reynolds number verification case, the trilinear scheme predicts the greatest error in settling velocity. Surprisingly however, the DGF method shows greater error than the \cite{Esmaily_Horwitz_2018} scheme throughout the bulk of the channel. This is likely owing to the fact that the DGF's obtained in section \ref{sec:find_dgf} were obtained at low Reynolds number, or equivalently $f/2\pi \rho  \nu^{2}\ll 1$. When the particle Reynolds number is $O(1)$, there is an additonal contribution to the flow disturbance generated by the point-force, owing to the advection of the disturbance velocity by itself via a term of the form $u'_j\frac{\partial u'_i}{\partial x_j}$, where $u'_{j}$ is the disturbance velocity. This non-linear term was negligible at low particle Reynolds number, but the results presented here, rather than suggesting a limitation of the DGF method as a whole, motivates caution of using a computational model outside the regime in which it was derived. The power of DGF is in fact that this non-linear advection term can be explictly incorporated into the construction of a DGF for this channel. The analysis in section \ref{sec:find_dgf} can be repeated at finite Reynolds number by computing the DGF for an Oseen-like system (see Appendix A for a derivation of an Oseen-DGF). The non-linear term can be factored into the analysis by linearizing $u'_j\frac{\partial u'_i}{\partial x_j} \approx u_j^{St}\frac{\partial u_i^{Os}}{\partial x_j}$. Note however, in problems with mean-advection, which is typical in most applications, the advection by the disturbance flow is typically small in comparison to advection of the disturbance by the mean flow, the latter which can be incorporated into the calculation of the DGF, without approximation (see Appendix A). 
We conclude this section by noting that the DGF method is the most accurate method in the near-wall region, where the effects of wall-damping override the inertial corrections to drag. In comparison, the \cite{Esmaily_Horwitz_2018} scheme predicts comparable levels of error in the near-wall region at finite particle Reynolds number, as predicted in the previous section at low particle Reynolds number.
%{\color{red} figure: finite Re verification}
\begin{figure}
  \centering
  \subfigure{\includegraphics[trim={3.5cm 7.6cm 3.5cm 8.2cm},clip, height=0.58\textwidth]{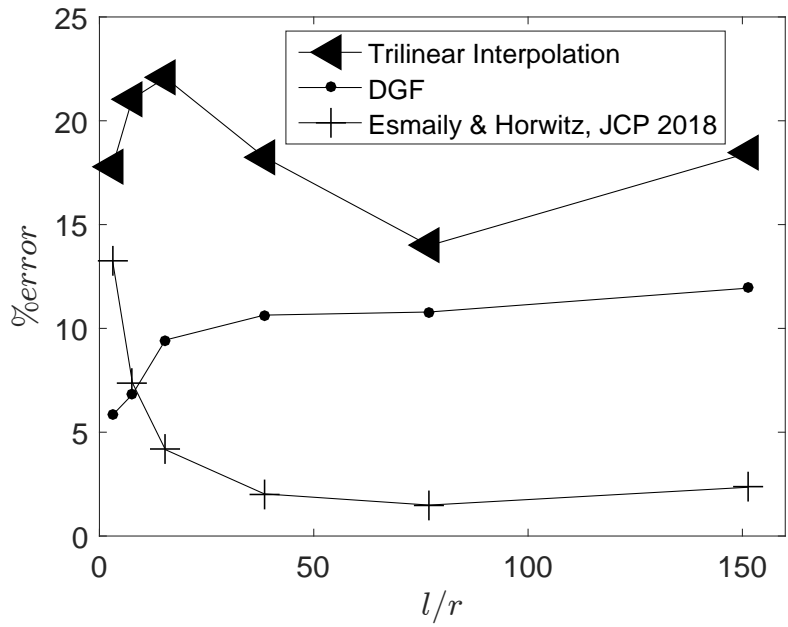} \label{fig:kf_100}}  
\caption{Percent error in settling velocity predicted by different schemes of a particle moving parallel to a plane wall for different wall-normal separations, $Re_{p} = 1$.}
\label{finiteRepverification}
\end{figure}
%\clearpage

}

\section{Application to particle-laden turbulent channel flow}
{
\label{sec:turbulent_channel}
As a final demonstration, we investigated a turbulent channel flow laden by particles to compare the effect of point-particle schemes on the predicted statistics. We use the same domain and grid spacing discussed in section \ref{sec:find_dgf}. We consider turbulence driven by a mean pressure-gradient whose friction Reynolds number in the absence of particles is $Re_{\tau}\approx 180$. The non-dimensional particle size, density ratio, and Stokes number are $d_p^{+}\approx 2.3, \rho_p/\rho_f = 175, St^{+} \approx 53$. The non-dimensional grid-spacing in wall-units is $\Delta x^{+}\approx 11.8, \Delta y^{+}\approx 0.4-5.8,\Delta z^{+}\approx 5.9$. Though the domain length and width are smaller than typical direct simulations of channel flow turbulence, we have verified that the mean and second order velocity profiles are in good agreement with the results of \cite{Kim_Moin_Moser_1987} under similar flow conditions (comparison curves not shown for clarity). We consider a mass-loading ratio of 0.51 which indicates a strong level of coupling between the particle and fluid phases. The particle coupling force is modelled using a Schiller-Naumann drag force, Eq.~(\ref{F_general_form}) without wall correction ($K(l/r)=1$). We compare the effect of trilinear, \cite{Esmaily_Horwitz_2018}, and DGF schemes on the predicted particle-laden fluid statistics.
%{\color{red} figure: rms components for three point-particle methods}
%\begin{figure}
%  \centering
%  \subfigure{\includegraphics[trim={3.5cm 8.0cm 3.5cm 7.5cm},clip, height=0.78\textwidth]{figures/comparison_high_mass_loading_revised_figure.pdf}
%  \put(-205,185){\includegraphics[trim={8.6cm 15.8cm 5.0cm 9.5cm},clip, width=60mm]{figures/legend_channel_flow.pdf}}
%   \label{fig:turbulent_channel}}  
%\caption{Profiles of normal components of rms Reynolds stress predicted by different Euler-Lagrange schemes in turbulent channel flow. Solid lines indicate stream-wise component, dotted indicate span-wise, and dashed indicate wall-normal component. Profiles have been normalized by wall friction velocity.} %The black reference curves are unladen channel flow data from \cite{Kim_Moin_Moser_1987} under similar particle-free conditions.}
%\end{figure}

\begin{figure}
  \centering
  \subfigure{\includegraphics[trim={3.5cm 8.0cm 3.5cm 7.5cm},clip, height=0.62\textwidth]{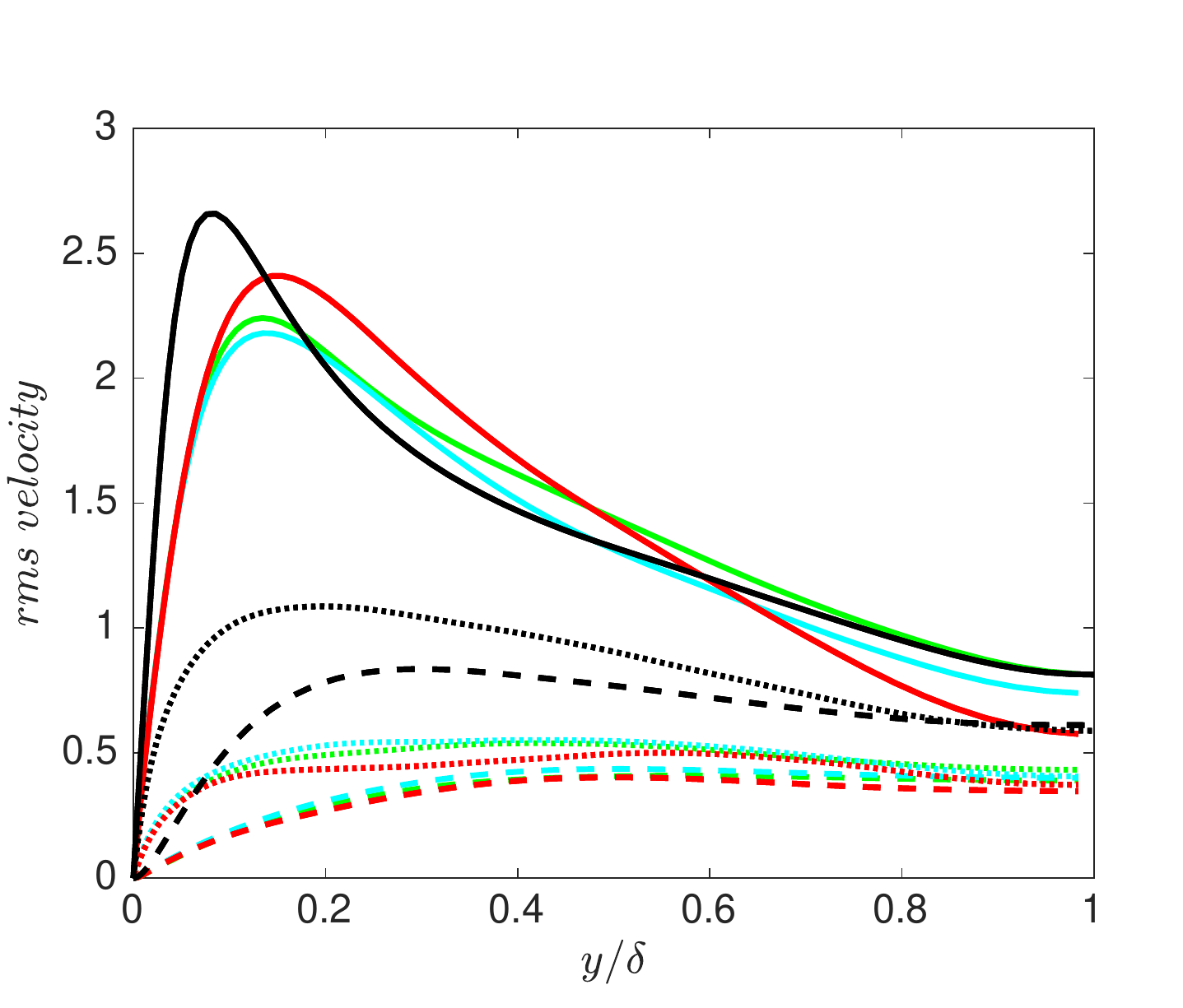}
  \put(-239,201){\includegraphics[trim={5.5cm 15.8cm 5.1cm 9.5cm},clip, width=74mm]{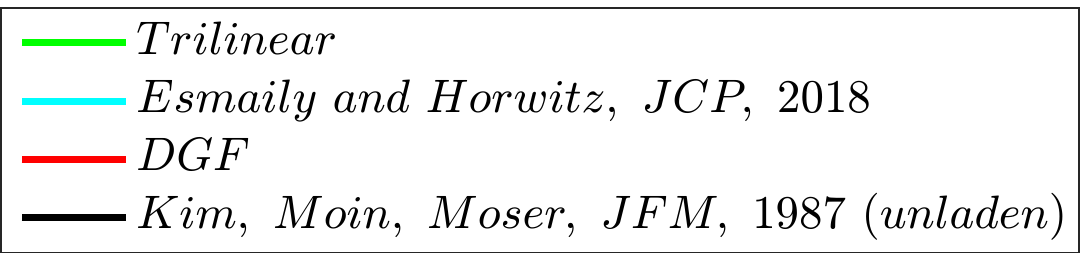}}
   \label{fig:turbulent_channel}}  
\caption{Profiles of normal components of rms Reynolds stress predicted by different Euler-Lagrange schemes in turbulent channel flow. Solid lines indicate stream-wise component, dotted indicate span-wise, and dashed indicate wall-normal component. Profiles have been normalized by wall friction velocity  {\color{black}of the unladen flow}. {\color{black}The black reference curves are unladen channel flow data from \cite{Kim_Moin_Moser_1987} under similar particle-free conditions.}}
\end{figure}

Wall-normal profiles of rms stream-wise, span-wise, and wall-normal fluctuations for different point-particle tracking schemes are shown in Figure~\ref{fig:turbulent_channel}. We also plot the unladen curves for comparison.
For all velocity components, it appears that the predictions of the trilinear and \cite{Esmaily_Horwitz_2018} schemes produce qualitatively similar predictions while the DGF prediction differs qualitatively in all three velocity statistics. The discrepancy is most notable in the stream-wise rms velocity profile. While all schemes predict the peak stream-wise fluctuations in the presence of particles to occur at approximately the same wall-normal location, the trilinear and \cite{Esmaily_Horwitz_2018} schemes predict nominal change in stream-wise velocity fluctuations near the center of the channel while the DGF method predicts noticable attenuation in the stream-wise rms component at the channel center. This observation may be explained by first noting that the particle Reynolds number profile (not shown), is order unity across the channel, so that the finite $Re_{p}$ verification problem considered in the previous section may be illustrative. Interestingly, the DGF and trilinear methods are less accurate at predicting the particle settling velocity (hence the coupling force) over the bulk of the channel compared with the \cite{Esmaily_Horwitz_2018} scheme. In contrast however, the DGF scheme is more accurate at predicting the coupling force in the near wall-region, while the trilinear and \cite{Esmaily_Horwitz_2018} schemes show similar levels of inaccuracy in the near-wall region. This near-wall region where the DGF approach is more accurate than the other methods, even at finite $Re_{p}$ corresponds to the viscous sub-layer and part of the buffer layer. This points to a mechanism of turbulence modification whereby particles trapped in the near wall region change turbulent dissipation which peaks in these inner-wall regions \citep{pope-2000}. 
We suspect that this modification to dissipation is transported from the wall to the channel center and accounts for the differences in the observed stream-wise velocity fluctuations at the channel center. Though less apparent, we note that the wall-normal and span-wise components predicted by the DGF method are also suppressed in comparison to the other schemes.

A more extensive comparison of these point-particle methods in particle-laden channel flow is underway and will be the subject of a future paper. Nevertheless, the results presented in this section, combined with the observations of the verification sections, highlight the importance of considering wall-damping on computation of the undisturbed fluid velocity and the fluid/particle statistics which are influenced by accurate/inaccurate computation of the point-force. %The accuracy of the DGF approach in the near-wall region 

}

\section{Conclusions}
\label{sec:conclusions}
We have presented a paradigm for Euler-Lagrange two-way coupled simulation involving the method of discrete Green's functions. This method is developed by exploiting linearity of the governing fluid equations. By considering the particular domain and discretization, the disturbance induced by a point-source can be recovered exactly. We derived the form of the discrete Green's function for the discrete Stokes equations and calculated the elements of this object for a discrete Stokesian response to a point-force applied at different points and directions between two fixed walls. These curves are the discrete analogues of the continuous analytical solutions obtained by \cite{Liron_Mochon_1975}. 

We then considered two verification problems involving the settling of a particle moving parallel to a plane boundary at different wall-normal separations. At low particle Reynolds number, consistent with the employed Stokesian DGF, the discrete Green's function approach developed in this work performed accurately at all wall-normal separations, compared with existing point-particle schemes which either were accurate in the bulk but not at the wall, or were not accurate at any wall-normal separations. At Reynolds number $O(1)$, the DGF approach remained the most accurate near the wall, but was not as accurate as the \cite{Esmaily_Horwitz_2018} scheme in the bulk of the channel. Rather than being a limitation of the DGF method, this points to the fact that the Stokesian DGF was being pushed to the edge of its applicability, where self-advection of the disturbance velocity becomes important. {\color{black}Though we have only obtained the Stokesian DGF for the present study,} in Appendix A we discuss a procedure for deriving a more accurate DGF at finite particle Reynolds number. The self-advection (non-linear term) can be handled by approximating the advection coefficient as a known function, e.g. $G^{1,1}$ and solving for the DGFs for the resulting linear problem. In a similar fashion, the \cite{Esmaily_Horwitz_2018} method was pushed to the edge of its applicability in the near-wall region, but recent work \citep{Pedram_APS_2019} suggests the method of \cite{Esmaily_Horwitz_2018} can be extended to the wall-bounded regime. These observations suggest that both DGF and extended \cite{Esmaily_Horwitz_2018} may become comparably accurate point-particle strategies at all wall-normal separations for turbulent channel flow simulation.

In the last section, we considered the application of the DGF method to the problem of particle-laden turbulent channel flow. When comparing the stream-wise velocity fluctuations predicted by the DGF method compared with other point-particle methods, it was found that {\color{black}the DGF method predicted considerable stream-wise velocity attenuation near the channel center, while the other approaches predicted small changes in stream-wise velocity at the center compared with the unladen flow.} {\color{black} Though requiring further study, we suspect} this observation is connected to the accuracy with which the coupling force, and hence the work-rate with which particles exert onto the fluid, can be applied to the fluid grid cells in the near-wall region where turbulence dissipation is highest. The DGF method is the most accurate method in this region. Collectively the observations in this section point to wall treatment as an important consideration in studying turbulence modification by particles. A more thorough comparison of point-particle methods will be the subject of future work. More particle-resolved simulations should also be conducted to assess the validity of the statistics predicted by different point-particle methods.

%Maxwell/temperature equations.
The Stokesian DGF obtained in this work can be extended to problems with mean-advection.  The coefficient of the advection term can be treated as a known mean-velocity profile plus $G^{1,1}$. Overall the non-linear self-advection term will be of second order effect compared with mean-advection in practical applications, so the higher error at finite Reynolds number can be remedied with Oseen-like DGF's (see Appendix A for details)\footnote{There has been recent work demonstrating that the method of Green's functions can be used to accurately approximate certain inhomogeneous nonlinear ordinary differential equations \citep{Frasca}. Though requiring more investigation, this suggests Landau-Squire type solutions \citep{batchelor-1967} could possibly serve as a basis for extending DGF to the full nonlinear Navier-Stokes equations or other nonlinear partial differential equations with source terms.}. The ability to be extended to finite Reynolds number makes DGF an attractive approach compared to the extended ERPP approach which is based on analytical solutions to the unsteady Stokes equations \citep{Battista_2019}. However, the ERPP approach may be more advantageous in situations where unsteady drag, e.g. history effects are important since an analytical solution in this case could be more favorable in comparison to storing time-dependent DGFs. Another potential advantage that DGF offers is the ability to estimate undisturbed quantities for PDEs and/or domains where analytical Green's functions owing to regularized point-source terms are unknown or cumbersome to express in closed form\footnote{Compare the Green's function of \cite{Blake_Chwang_1973} to that of \cite{Liron_Mochon_1975} which should be similarly reliable for a small particle near a single plane wall but the latter work which accounts for the presence of a second wall is expressed in a comparatively complicated form.}.

This work, besides providing a practical approach to estimate undisturbed quantities in Euler-Lagrange simulation, provides insight into the intimate connection among interpolation/projection stencil (Lagrange polynomial vs. Gaussian e.g.) for Euler-Lagrange transfer, governing fluid equations and their numerical discretization (2nd vs. 4th order finite difference e.g.)/element structure (structured vs. unstructured e.g.) and boundary conditions (periodic vs. wall e.g.) and the ultimate discrete response generated by discrete source terms. %{\color{red}Awkward wording (discrete/continuous and solution twice): This paradigm should be recognized %as not just a method for computing undisturbed quantities, but should be recognized 
%as a fundamental connection between continuous (discrete??) solution of PDEs with regularized impulsive source terms and their corresponding discrete solutions.} 
While we have applied this method to hydrodynamic equations, we expect this approach would be of value in the consideration of thermal and electricity/magnetism effects experienced by particles. The Stokesian DGFs obtained for this work may be of value to other researchers conducting investigations of particle-laden channel flow and we can make these DGFs available upon request.

%We calculated the discrete Green's functions of the Stokes equations in a two-plane geometry. These curves are the discrete analogues of the continuous analytical solutions obtained by \cite{Liron_Mochon_1975}.

%response to criticism of finite Rep results, maybe this discussion goes in the Appendix

\section*{Acknowledgments}
%Some of the results in this work were presented at the 2016 American Physical 
%Society Division of Fluid Dynamics meeting.
Funding for this work was provided through a
United States Department of Energy grant for the Predictive Science Academic 
Alliance Program 2 (PSAAP2) at Stanford University under grant number  DE-NA0002373.
This work was performed under the auspices of the U.S. Department of Energy by Lawrence Livermore 
National Laboratory under contract DE-AC52-07NA27344. LLNL-JRNL-794560.

\section*{Declaration of Interests}
The authors report no conflict of interest.

%\clearpage

\section*{Appendix A: Extension of discrete Green's functions to finite Reynolds number}
{
\label{sec:appendix}
%{\color{red}need to complete this derivation}
Here we consider a finite Reynolds number extension to the Stokes discrete Green's function. Here we will consider a base fluid flow $\mathbf{\bar{u}}=\big(\bar{U}(y),0,0\big)$. Let us denote the perturbation induced by the two-way coupling force as $\mathbf{u'}$. Starting with the steady Navier-Stokes equation and writing the fluid velocity as $\mathbf{u}= \mathbf{\bar{u}}+\mathbf{u'}$, we have:

\begin{equation}
\label{Oseen_1}
\rho_f u_j\frac{\partial u_i}{\partial x_j}=\mu\nabla^2u_i-\frac{\partial p}{\partial x_i}- F_{i} 
\end{equation}

\begin{equation}
\label{Oseen_2}
\nabla^2 p = -\rho_f \frac{\partial u_j}{\partial x_i}\frac{\partial u_i}{\partial x_j} -\frac{\partial F_i}{\partial x_i}
\end{equation}

The advection term, $u_j\frac{\partial u_i}{\partial x_j}$ can be decomposed as:

\begin{equation}
\label{decomposition}
u_j\frac{\partial u_i}{\partial x_j}=\bar{u}_j\frac{\partial \bar{u}_i}{\partial x_j}+\bar{u}_j\frac{\partial u'_i}{\partial x_j}+u'_j\frac{\partial \bar{u}_i}{\partial x_j}+u'_j\frac{\partial u'_i}{\partial x_j}
\end{equation}

The first term in (\ref{decomposition}) is identically zero. The remaining terms can be written, along with their associated scalings, as:

\begin{equation}
\label{reduced_decomposition_1}
\bar{u}_j\frac{\partial u'_i}{\partial x_j} = \bar{U}\big(\frac{\partial u'}{\partial x},\frac{\partial v'}{\partial x},\frac{\partial w'}{\partial x}\big)\sim \frac{u'\bar{U}}{\Delta}
\end{equation}

\begin{equation}
\label{reduced_decomposition_2}
u'_j\frac{\partial \bar{u}_i}{\partial x_j}=\big(v'\frac{\partial \bar{u}}{\partial y},0,0\big)\sim \frac{u'\bar{U}}{\delta}\footnote{The proper scaling near the wall, $\sim \frac{u'\bar{U}}{\Delta}$ is hereby noted, and the current analysis can be analagously extended to include this term.}
\end{equation}

\begin{equation}
\label{reduced_decomposition_3}
u'_j\frac{\partial u'_i}{\partial x_j} \sim \frac{u'^2}{\Delta}
\end{equation}

Using typical channel dimensions, grid-spacing, and disturbance velocity scalings,

\begin{equation}
\bar{u}_j\frac{\partial u'_i}{\partial x_j}\gg \Bigg\{u'_j\frac{\partial \bar{u}_i}{\partial x_j}, u'_j\frac{\partial u'_i}{\partial x_j} \Bigg\}
\end{equation}

Therefore we seek the discrete Green's function associated with the reduced discrete linearized Oseen system:

\begin{equation}
\label{Oseen_1_discrete}
\rho_f \bar{u}^n_j\frac{\delta u^n_i}{\delta x_j} = \mu\frac{\delta^2u_i^n}{\delta x_j \delta x_j}-\frac{\delta p^n}{\delta x_i}- F_{i}^n 
\end{equation}

\begin{equation}
\label{Oseen_2_discrete}
\frac{\delta^2 p^n}{\delta x_j \delta x_j} = -\rho_f \frac{\delta \bar{u}^n_j}{\delta x_i}\frac{\delta u^n_i}{\delta x_j} -\frac{\delta F_i^n}{\delta x_i}
\end{equation}

Defining the following discrete differential operators: $C^{kn} \equiv (\nabla^2)^{-1}$, $A_j()_j\equiv \frac{\partial }{\partial x_j}()_j$, and $B_{i} =\mathbf{\nabla}$, and the quantity, $D^{n}_i = \rho_f\bar{u}_j^n B_j u_i^n$ with no summation of the upper indices, we can write the formal solution to Eqs. (\ref{Oseen_1_discrete}) and (\ref{Oseen_2_discrete}) as:

\begin{equation}
\label{discrete_pressure_Oseen}
p^k = -C^{kn}\Big\{\ A_i D^{n}_i  + A_jF_j^n \Big\}
\end{equation}

%\begin{equation}
%\label{discrete_velocity_Oseen}
%u_i^n = \mu^{-1}C^{nk}(\delta_{ij}^{kp}-B_iC^{kp}A_j)F_j^p
%\end{equation}

%{\color{red}Here I showed more algebra to help check I didn't make a mistake, we can reduce this later if we decide to.}

\begin{equation}
\label{Oseen_3_discrete}
\big\{\delta^{kp}_{ij}- B_iC^{kp} A_j\big\} D^{p}_j - \mu A_jB_ju_i^k = \big\{ B_iC^{kp} A_j - \delta_{ij}^{kp}\big\}F_j^p
\end{equation}

\begin{equation}
\label{Oseen_4_discrete}
\big\{\delta^{kp}_{ij}- B_iC^{kp} A_j\big\} \rho_f\bar{u}_m^{(p)} B_m u_j^p - \mu A_jB_ju_i^k = \big\{ B_iC^{kp} A_j - \delta_{ij}^{kp}\big\}F_j^p\footnote{the (p) superscript is used to emphasize that while the velocity is evaluated at grid point p, there is no summation of index p}
\end{equation}

\begin{equation}
\label{Oseen_5_discrete}
\big\{\delta^{kp}_{ij}- B_iC^{kp} A_j\big\} \rho_f\bar{u}_m^{(p)} B_m u_j^p - \mu A_mB_m\delta_{ij}^{kp} u_j^p = \big\{ B_iC^{kp} A_j - \delta_{ij}^{kp}\big\}F_j^p
\end{equation}

\begin{equation}
\label{Oseen_6_discrete}
\Big\{ \big\{\delta^{kp}_{ij}- B_iC^{kp} A_j\big\} \rho_f\bar{u}_m^{(p)} B_m  - \mu A_mB_m\delta_{ij}^{kp} \Big\}u_j^p = \big\{B_iC^{kp} A_j - \delta_{ij}^{kp}\big\}F_j^p
\end{equation}

By letting $E^{nk} = \Big\{ \big\{\delta^{kp}_{ij}- B_iC^{kp} A_j\big\} \rho_f\bar{u}_m^{(p)} B_m  - \mu A_mB_m\delta_{ij}^{kp} \Big\}^{-1}$ and

 $G_{ij}^{np} = E^{nk}\big\{B_iC^{kp} A_j - \delta_{ij}^{kp}\big\})$, Eq. (\ref{Oseen_6_discrete}) can be written in a simpler form:

\begin{equation}
\label{DGF_Oseen_equation}
u_i^n = G_{ij}^{np}F_j^p
\end{equation}
}

\section*{Appendix B: Staggered grid considerations}
{
\label{sec:appendixB}

\begin{figure}
  \centering
  \subfigure{\includegraphics[trim={5.5cm 0.5cm 3.5cm 0.5cm},clip, height=0.81\textwidth,page=3]{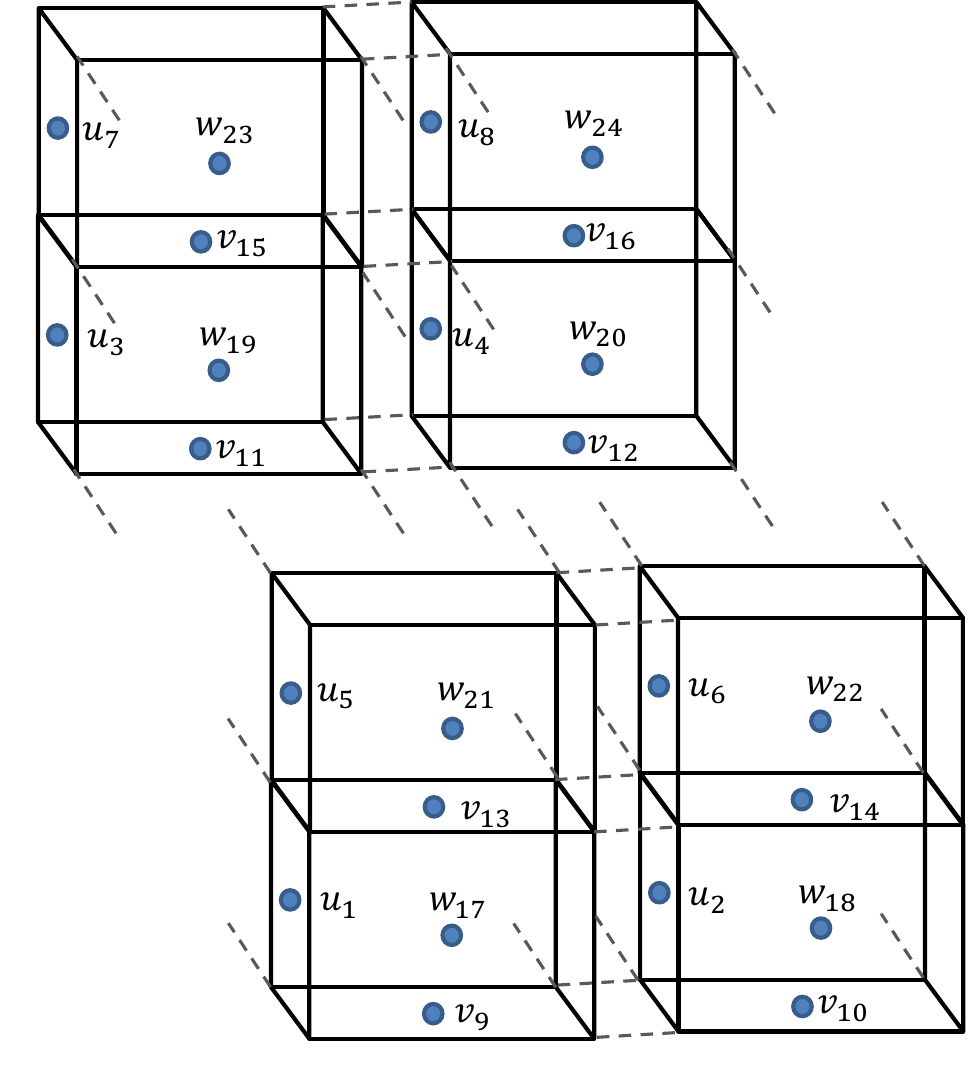} \label{fig:24_gridpoints_staggered}}  
\caption{Gridpoints involved in trilinear interpolation and projection {\color{black}for the staggered arrangement considered in this work}. For an arbitrary location of a particle in the domain, the surrounding gridpoints can be mapped to the notation used here. Streamwise cellfaces are denoted with $u$, spanwise with $w$, and wall-normal with $v$.}
\label{fig:24_staggered}  
\end{figure}

{\color{black}The staggered grid incorporated in this work, shown in Figure~\ref{fig:24_staggered} in general requires more storage per wall-normal location than the $24\times 24$ elements required for a collocated configuration. To illustrate this, consider a particle traveling exclusively in the stream-wise direction from the x-coordinate of $w_{17}$ to the x-coordinate of $w_{18}$. The initial y-coordinate is the same as that of $u_{1}$ and the initial z-coordinate is half-way between the z-coordinate of $u_{1}$ and $u_{3}$. When the particle passes the x-coordinate of $u_{2}$, interpolation and projection of Lagrangian information will now require information from $u$-cell faces downstream, outside of the set $u_{1-8}$, even though the gridpoints associated with span-wise and wall-normal interpolation remain unchanged. In other words, Figure~\ref{fig:24_staggered} depicts one of eight needed interpolation stencils for a trilinear configuration. This presents no issue for the diagonal blocks of the DGF elements since, for example the $G^{25,25} = G^{1,1}$, where $25 $ is the down-stream stream-wise cell-face considered in the previous example. Similarly $G^{1,2}=G^{2,25}$ etc. However, the issue is that the element $G^{21,25}$ for example, in other words an off-diagonal-block element is not equal to any of the $24\times 24$ stored elements. In general, this means a staggered grid implementation of DGF requires more elements than the collocated grid counterpart. To reduce storage cost however, we have chosen to store the DGF elements associated with the canonical arrangement shown in Figure~\ref{fig:24_staggered}. This means that when using Eq.~(\ref{estimate_undisturbed}) to estimate the undisturbed fluid velocity vector, $W^{k}$ and $w^{n}$ will provide the correct interpolation weights but certain off-diagonal block elements of the DGF matrix outside the canonical configuration will be approximated based on DGF elements obtained in the canonical configuration. This approximation will lead to negligible error %e reason this can be done 
because off-diagonal blocks constitute a small correction to that contributed by diagonal blocks of the DGF matrix. In other words, for a three dimensional force, most of the x-component of the response e.g. is contributed by the x-component of the force projected to stream-wise cell-faces. This is readily apparent in Figure~\ref{fig:contours}. The elements of the diagonal blocks are clearly much greater than the elements of the off-diagonal blocks. The exception is the wall-normal block near the wall (Figure~\ref{fig:contour_wall}), because a wall-normal response cannot be generated close to the wall. %To quantify the effect of the off-diagonal blocks, we assessed the $L^{2}$ norm of the off-diagonal blocks and found this value to be two orders of magnitude lower than the $L^{2}$ norm of the diagonal blocks. 
To quantify the importance of the diagonal blocks, the square of the Frobenius norm of the whole DGF-matrix is compared with the sum of the squares of the Frobenius norms of the diagonal blocks, that is $(|G^{(1:8,1:8)}|^{2}+|G^{(9:16,9:16)}|^{2}+|G^{(17:24,17:24)}|^{2})/|G^{(1:24,1:24)}|^{2}$. The respective ratios for the near-wall and near center DGF matrices are approximately $0.97$ and $0.95$, thus demonstrating dominance of the diagonal block elements. The success of the present implementation of a Stokesian DGF is corroborated by the low Reynolds number verification study shown in section~\ref{sec:lowRepverifcation}.
%To quantify the importance of the diagonal blocks, the square of the Frobenius norm of the whole DGF-matrix is compared with the sum of the squares of the Frobenius norms of the diagonal blocks, that is $(|G^{(1:8,1:8)}|^{2}+|G^{(9:16,9:16)}|^{2}+|G^{(17:24,17:24)}|^{2})/|G^{(1:24,1:24)}|^{2}$. The respective ratios for the near-wall and near center DGF matrices are approximately $0.97$ and $0.95$, thus demonstrating dominance of the diagonal block elements. 

}
}
%\clearpage
\bibliography{MS}
\bibliographystyle{unsrt}

\end{document}